\documentclass[a4paper,11pt]{article}
\usepackage{enumitem}

\usepackage[a4paper,
  left=2.5cm, right=2.5cm,
  top= 3cm, bottom=4cm]{geometry}
\usepackage{amsmath}
\usepackage{oldgerm}
\usepackage{amssymb}
\usepackage{bbm}
\usepackage[dvips]{graphicx}
\usepackage{epsfig}
\usepackage{color} 
\usepackage{cite}
\usepackage{epic}
\usepackage{hyperref}

\usepackage{empheq}

\numberwithin{equation}{section}

\usepackage[utf8]{inputenc}

\usepackage{graphicx}
\usepackage{mathrsfs}
\usepackage{amsmath}
\usepackage{amssymb}
\usepackage{braket} 
\usepackage{hyperref}
\usepackage{frontespizio}
\usepackage{comment}
\usepackage{epsfig}
\usepackage{float}
\usepackage{color}
\usepackage[title]{appendix}
\usepackage{multirow}
\usepackage{tikz}
\usetikzlibrary{decorations.pathmorphing}
\usepackage{varwidth}
\usepackage{tikz}
\usetikzlibrary{backgrounds,patterns,calc}
\usepackage{xparse}
\usepackage{caption}
\usepackage{enumitem}
\setlist[itemize]{leftmargin=*}

\usetikzlibrary{decorations.pathmorphing}

\tikzset{zigzag/.style={decorate, decoration=zigzag}}

\newcommand{\citep}{\cite}

\newcommand{\bea}{\begin{eqnarray}}
\newcommand{\eea}{\end{eqnarray}}
\newcommand{\be}{\begin{equation}}
\newcommand{\ee}{\end{equation}}
\newcommand{\ba}{\begin{align}}
\newcommand{\ea}{\end{align}}
\newcommand{\exd}{{\rm d}}

\def\ssL{{\scriptscriptstyle L}}
\def\ssF{{\scriptscriptstyle F}}
\def\ssI{{\scriptscriptstyle I}}
\def\ssN{{\scriptscriptstyle N}}
\def\ssR{{\scriptscriptstyle R}}

\def\ssW{{\scriptscriptstyle W}}
\def\ssY{{\scriptscriptstyle Y}}
\def\EM{{\scriptscriptstyle EM}}
\def\SM{{\scriptscriptstyle SM}}
\def\PQ{{\scriptscriptstyle PQ}}

\usepackage{subfigure}

\def\0{{\boldsymbol 0}}

\def\cL{\mathcal L}

\title{}
\author{}

\numberwithin{equation}{section}

\begin{document}
%
%
\begin{titlepage}


\vspace{-0.5cm} {\flushright {\small{.}}} \\

%
\begin{center}

\vspace{1.2truecm}

{\huge\bf{
Steven Weinberg: A Scientific Life\\[0.3cm]}}

\vspace{1.2truecm}

{\fontsize{10.5}{18}\selectfont
{\bf  C.P. Burgess$^{a,b,c}$ and F. Quevedo$^{d,e}$
}}
\vspace{.5truecm}

{\it{\footnotesize
${}^a$ Department of Physics \& Astronomy, McMaster University, 
1280 Main Street West, Hamilton ON, Canada.\\
${}^b$ Perimeter Institute for Theoretical Physics, 
 31 Caroline Street North, Waterloo ON, Canada.\\
${}^c$ School of Theoretical Physics, Dublin Institute for Advanced Studies,
10 Burlington Road,  Dublin, Ireland.\\
${}^d$ DAMTP, University of Cambridge, Wilberforce Road,  Cambridge, 
United Kingdom.\\
\vspace{-1mm}
${}^e$ New York University Abu Dhabi, 
Saadiyat Island, Abu Dhabi, United Arab Emirates.
}}



  \vskip 2.2cm

 \begin{abstract}
{\small{{Steven Weinberg was a giant of late 20th Century physics on whose shoulders we stand while groping for the science of the 21st Century. This article provides a too-brief summary of a selection of his many achievements -- eight decades of superlative research, eight classic textbooks, eight best-selling forays into popular science writing and more.\footnote{An abridged version of this article will appear in Biographical Memoirs of Fellows of the Royal Society.}
 }}}
 \end{abstract}
\end{center}

\end{titlepage}







\renewcommand*{\thefootnote}{\arabic{footnote}}
\setcounter{footnote}{0}

\newpage

\tableofcontents

\section{Introduction}

When Steven Weinberg started his scientific journey he felt himself to be at the margins of the intellectual currents of the time. For much of the community the problems of divergences seemed to show that quantum field theories (QFTs) -- the topic on which he largely worked -- were a dead end. By the time of his passing on July 23, 2021 his point of view had become the \emph{Zeitgeist} of the field and Steven Weinberg was widely regarded as its premier scientist. If the measure of an intellect is the influence it has on those that surround it, then at his passing Steven Weinberg was arguably the greatest theoretical physicist alive. 

This major claim can be justified on many grounds. One of these is the work for which he received his Nobel prize in 1979: the unification of the electromagnetic and weak interactions. It falls only to a handful of scientists to identify a common thread in what had previously been regarded as different types of interactions and thereby unify them into a single comprehensive framework. Isaac Newton was perhaps the first to do so with the realization that the gravitational force we experience on Earth is the same as the force that guides planets and moons in their orbits within the solar system. James Clerk Maxwell took a similar leap by recognizing that the electric and magnetic forces are just different glimpses into the phenomenon of electromagnetism. Albert Einstein joined this august group by identifying the physics of gravity with the physical rules governing the previously passive geometry of space and time within which physics is framed. The electroweak unification of Glashow, Weinberg and Salam lies in this same category of achievement. 

Remarkable as this was, the discovery of the electroweak theory can be argued to be not his most important scientific legacy. Steven Weinberg also changed what it is that scientists think they are doing when they discover new theories like these. Previous discoverers of fundamental physical laws regarded themselves as identifying the pristine rules governing how Nature works. But Weinberg taught us that what one really does is to uncover the outermost layer of the onion of knowledge; we do not identify fundamental truths but instead find effective descriptions of Nature that apply at low energies but generally break down at higher energies (or smaller distances). Although this kind of picture was not in itself new, what was new was the development of tools -- Effective Field Theories (EFTs) -- that systematized how to extract information one scale at a time. The great conceptual simplicity of these tools allow predictions to be made where they were not previously possible, causing them to be adopted across many fields beyond the ones in which Weinberg started. They also have pointed a fruitful way forward in long-standing problems like reconciling quantum mechanics with gravity, whose ultimate resolution is not yet settled.  

Besides having an extremely prolific research career extending over eight decades, he also spoke to a much wider audience. Countless new students and mature researchers learned and honed their trade reading his eight classic textbooks. These are widely acknowledged as standard references in each of several different fields, putting the lie to the witticism that those who cannot do, teach. His award-winning popular science books were bestsellers; engagingly explaining the intricacies of cosmology and elementary particle physics to lay audiences. They have since been translated into many languages and inspired generations of scientists -- and no doubt will continue to do so for generations to come. 

As his success and fame grew he became a public figure and throughout his long career used this platform to advocate on behalf of science, justice and critical thinking. This has spawned an often controversial dialogue with social scientists and others about the very nature of scientific knowledge and of the role of religion in society. His essays on these subjects reveal a refreshing common sense laced with a sharp wit and a gift for the pithy phrase. One such is his view expressed about the relationship between religion and morality \cite{DesignerU}:
\begin{quote}
{\it With or without religion, good people can behave well and bad people can do evil; but for good people to do evil — that takes religion.}
\end{quote}

As a scholar and writer, he is the epitome of the ‘Big Picture’ person who moves among the very few truly Renaissance Thinkers.

Being such a prominent figure means there exists much material online that documents his achievements, including a forthcoming autobiography \cite{autobiography}. Here we hope to provide a scientific perspective, briefly summarizing a selection of his scientific contributions that we believe provide glimpses into his spectacular scientific successes.\footnote{Interested readers should also consult Michael Duff's excellent collection of his papers with personal comments and insights into the relevance of his work \cite{duff}.} When detailing his popular books we rely on liberal quotes to allow him speak with his own voice.  

\subsection{Brief Biography}

Steven Weinberg was born in New York City on May 3, 1933, the son of Jewish immigrants Frederick Weinberg and Eva Israel. He attended school in Manhattan and  became interested in science at a young age encouraged by his father. At age 15-16 he became interested in theoretical physics at the Bronx High School of Science, where together with a few classmates -- including Sheldon Glashow, with whom he would share the Nobel prize in 1979 (and with Abdus Salam \cite{nobel}) -- he became active on several scientific initiatives, including the creation of a science fiction zine and narrating a production of Orson Well's `War of the Worlds' \cite{glashow}. 

His post-secondary career began in 1950. Although rejected by Harvard, he was accepted into other top universities and of these he, Glashow and other friends decided to enroll in Cornell University's physics program. There he joined the honorary fraternity, Telluride House, and met his future wife, the lawyer Louise Goldwasser. They married in 1954 before graduating from Cornell the same year. Weinberg spent the next year at the Niels Bohr Institute and then moved to Princeton where he completed his PhD  in 1957 under the supervision of Sam Treiman. His thesis title was `The role of strong interactions in decay processes'. 

After graduation he became a junior researcher at Columbia University. In 1959, after being turned down for a faculty position there, he moved to the University of California at  Berkeley, first as a research fellow and then as faculty from 1960 to 1966. It was there that his daughter Elizabeth was born in 1963.  In 1966 he became a lecturer at  Harvard, the next year he went to MIT as a visiting professor and from 1969 as a full  professor. In 1973  he moved back to Harvard as the Higgins  Professor of Physics, replacing Julian Schwinger after his retirement. He was also appointed as Senior Scientist at the Smithsonian Astrophysical Observatory. In 1982 he moved to the University of Texas at Austin as the Josey Regental Professor of Science, where he stayed until his death. His wife Louise Weinberg remains there as a law professor.

\section{Research Highlights}

In so prolific a career only a few highlights can be touched on. Although most would agree on what his biggest achievements were, even his lesser-known work often proved to be foundational. We provide here our personal collection of favourites for each stage of his life.

\subsection{Early years}
During his time at Columbia, Weinberg wrote several phenomenological papers with different collaborators, including Robert Marshak, George Sudarshan and Sam Treiman, but his main work -- as became standard throughout his career -- was published as single-author papers. Two of those written in the 1950s stand out as `renowned' articles ({\it i.e}~articles receiving over 500 citations in the InSPIRE High Energy Physics \href{https://inspirehep.net}{database}) that are still being referred to after more than 60 years. 

The first of these \cite{Weinberg:1958ut} establishes general results on the recently discovered $V$--$A$ \cite{V-A}
 version of the Fermi theory of weak interactions \cite{Fermi}. Using general symmetry arguments he established two classes of weak interactions depending on their transformation properties under a combination of charge conjugation and an isospin rotation (called $G$-symmetry) that remains useful even now when studying hadronic weak interactions. 

His second noteworthy paper of this time \cite{Weinberg:1959nj} proves a very useful result (`Weinberg's theorem') about quantum field theory. In it he identifies the asymptotic behaviour of Green's functions at high energies and uses this to explore the ultraviolet (UV) convergence of scattering amplitudes. This work closed the last loophole in the  perturbative renormalization program for theories like Quantum Electrodynamics (QED), and thereby put previous work by Freeman Dyson \cite{Dyson:1952tj} and others onto a firm foundation. This work also brought him into contact with the research of Abdus Salam and -- as he mentioned on several occasions -- began a career-long history of mutual respect that included future collaborations and sharing the Nobel prize. 

\subsection{1960s}

The 1960s were arguably Weinberg's most fruitful decade of research, in which he laid the foundation for much of his later impact on the direction of physics. In retrospect this was the decade where he chose an intellectual road that was (at that time) less travelled; a road that now is one of science's central thoroughfares. Unlike many at that time, he largely dedicated himself to the study of QFT and its phenomenological implications. In these studies his signature style also emerges: a search for the most general mathematically sound way to approach a problem chosen for its potential physical implications.

\begin{itemize}

\item {\bf Spontaneously broken symmetries}

In 1962 Weinberg visited Salam at Imperial College in London, where they wrote -- with Jeffrey Goldstone -- a classic paper on quantum field theory entitled `Broken Symmetries' \cite{Goldstone:1962es}. This paper is now an obligatory read in many areas of theoretical physics and reveals the essence of Weinberg's bulldozer-like approach to solving physics problems: proving in full generality an important physical result -- in  this case a result that Goldstone and Yoichiro Nambu had previously highlighted in particular examples.

The paper offers three independent proofs of what is now known as Goldstone's theorem, which states that every system whose ground state is not invariant under a continuous global symmetry (a `spontaneously broken' symmetry) must have a gapless state whose properties and low-energy interactions are universal, in the sense that they are largely determined by the symmetry that is broken. For most applications (the exception being spontaneously broken global supersymmetry) these states are bosons and so are usually called Goldstone bosons. In the simplest weakly coupled relativistic examples these bosons are massless scalars that emerge as particular eigenstates when a theory's scalar potential has minima that are not invariant under symmetries. But the generality of the proof in this paper shows that their existence does not rely on weak coupling or the kinematics of Special Relativity.

 \item{\bf Chiral perturbation theory} 
 
Weinberg's interest in spontaneously broken symmetries was driven by the desire to use their physical implications, to start with in the strong interactions where symmetry arguments were likely to be particularly useful given that the strength of the interactions precluded using standard perturbative methods. On the advice of Julian Schwinger\footnote{See \cite{Weinberg:2016kyd} for a detailed description of this story in his own words.} and building on ideas of current algebras and PCAC (`Partially Conserved Axial Currents' -- a dynamical assumption that essentially implies that the observed strongly interacting pions were Goldstone bosons for spontaneously broken approximate symmetries), he exploited the universality of Goldstone boson interactions to derive what are now known as `soft-pion' theorems \cite{Weinberg:1966kf}. These are universal expressions for the form taken by pion scattering amplitudes at low energies that follow purely from their interpretation as Goldstone bosons for an approximate global chiral $SU_\ssL(2) \times SU_\ssR(2)$ symmetry of which only the diagonal (isospin) subgroup $SU_\ssI(2)$ preserves the ground state.\footnote{We now know that these symmetries can be regarded as acting on the quarks from which the observed strongly interacting hadrons are built, with the left- and right-handed up and down quarks transforming independently as doublets.} 

Subsequent progress came with the realization that the universality of the soft-pion theorems allows them to be derived directly from a lagrangian built using elementary pion fields, provided only that this lagrangian is designed to involve the fewest possible derivatives and to incorporate the $SU_\ssL(2) \times SU_\ssR(2) \to SU_\ssI(2)$ symmetry-breaking pattern \cite{Weinberg:1968de}. In this lagrangian the spontaneously broken symmetries are distinguished because their action on the pion fields are nonlinear (unconventional for symmetries examined at the time), and this is all that is needed to characterize their low-energy interactions. 

The first key observation was that there is no loss of generality in the assumption of using a quantum field theory with elementary pion fields to realize the symmetry,\footnote{This argument is made explicit in a lovely subsequent essay \emph{What is quantum field theory and what did we think it is?} \cite{Weinberg:1996kw}.} even though the pions are not themselves elementary (in the modern picture they are built from constituents like quarks and gluons). The second key observation was that this elementary scalar lagrangian can be treated with standard perturbative methods (again despite the underlying interaction involving strong couplings) because Goldstone's theorem also ensures Goldstone bosons interact only weakly at low energies compared to the symmetry-breaking scale. The replacement of strongly coupled dynamics with emergent weakly interacting Goldstone fields makes the result arguably the first premeditated use of EFT methods in the modern sense of the term. This framework was subsequently generalized to an arbitrary symmetry breaking pattern $G \to H$ by Coleman, Callan, Wess and Zumino \cite{Coleman:1969sm, Callan:1969sn}, and in this form provides the modern understanding of Goldstone boson dynamics. 

This work was completed at a busy time: his main result in this direction was sent to publication on 25 September 1967, less than a month before his groundbreaking article on electroweak unification (more about which below). 

 

\item{\bf Soft theorems and gauge theories} 

Another highlight of this epoque is Weinberg's study of the foundations of quantum field theory, providing the general solution to the problem of identifying which fields can represent which types of particles (for arbitrary spins) \cite{Weinberg:1964cn, Weinberg:1964ev, Weinberg:1969di}. This leads to the observation that the `gauge principle' (the assumption that massless -- or parametrically light -- spin-1 particles couple to one another and to matter in a gauge-invariant way) is not actually a separate assumption, independent of the usual requirements of Lorentz invariance and unitarity (say) in quantum mechanics. Gauge invariance is better thought of as a consequence of combining quantum mechanics with Special Relativity in the low-energy limit, and a similar conclusion leads to general covariance for massless spin-2 particles \cite{Weinberg:1964ew, Weinberg:1965rz, Weinberg:1965nx}. 

For massless spin-1 particles the essence of the argument relies on the observation that the polarization vector  $\varepsilon^\mu(k)$ for massless spin-1 particle of 4-momentum $k^\mu$ is \emph{not} actually a 4-vector. It only transforms as a 4-vector under Lorentz transformations \emph{up to a gauge transformation} (that is, $\varepsilon^\mu(k)$ acquires a shift proportional to $k^\mu$ in addition to the usual 4-vector transformation rule. As a consequence the leading low-energy interactions involving massless photons can only be Lorentz invariant if they are also gauge invariant. 

Similar reasoning also leads to new types of gauge symmetries for massless particles with spins higher than 1, with general covariance emerging for the dominant low-energy couplings of any massless spin-2 particle. Later extensions of his argument show that supersymmetry is the gauge invariance implied by massless spin-$\frac32$ particles. For massless particles with spins higher than 2 the required symmetry would have to not commute with Lorentz transformations, a possibility that is excluded in the presence of interactions -- apart from supersymmetry and general covariance -- in relativistic theories by the Coleman-Mandula theorem \cite{Coleman:1967ad} and its supersymmetric extension \cite{Haag:1974qh}. Nature requires interacting massless particles to have helicity $\leq 2$; an extraordinarily powerful result.

As was his wont, Weinberg's arguments were both practical and general. As part of this cluster of arguments he derives general expressions for the amplitude obtained when a low-momentum (soft) photon is emitted by one of the particles involved in some scattering process and uses these `soft theorems' to study the infrared behaviour of photon and graviton emission and exchange processes \cite{Weinberg:1965rz}. Unlike for Goldstone bosons the interactions of gauge particles are \emph{not} necessarily weak at low energies, leading for photons to scattering amplitudes that diverge in the infrared (\emph{i.e.}~when summed over very low-energy photon exchange). Weinberg's treatment provides a particularly lucid description of how these quantum divergences in electromagnetism intervene to solve the famous `infrared catastrophe' of classical electromagnetism (or, equivalently, provides a proof of the Bloch-Nordsieck theorem \cite{Bloch:1937pw} that shows how infrared divergences in Quantum Electrodynamics are cancelled by divergences in the multiple emission of soft photons).

These calculations also make the emergence of gauge symmetries very concrete. Demanding soft-photon amplitudes to be invariant under Lorentz transformations (and so also under gauge transformations) requires the sum of the incoming charges to equal to the sum of the outgoing charges:
\begin{equation}
\sum_{\rm incoming}Q_i=\sum_{\rm outgoing}Q_i \,,
\end{equation}
showing concretely how conservation of electric charge can be derived directly from general principles of Lorentz invariance. The same argument applied to soft graviton emission instead implies the condition
\begin{equation} \label{momcons}
\sum_{\rm incoming} \kappa_i p_i^\mu =\sum_{\rm outgoing} \kappa_i p_i^\mu
\end{equation}
where $p_i^\mu$ are the components of 4-momentum for the $i$-th particle and $\kappa_i$ is the corresponding `gravitational charge'. This linear relation among momenta can be satisfied only if the corresponding charges are all equal -- \emph{i.e.}~$\kappa_i=\kappa$ is independent of particle species. In this case (\ref{momcons}) reproduces the well-known condition of momentum conservation while the condition $\kappa_i = \kappa$ shows that general principles require gravitational interactions to be universal (a quantum derivation of the equivalence principle). Extending this to higher spins gives quadratic and higher order relations amongst the momenta that cannot be satisfied kinematically, more concretely showing why interacting massless particles cannot have helicity larger than 2. 

In these papers Weinberg further shows that the low-energy propagation of photons and gravitons is necessarily described by the linearised version of both Maxwell's and Einstein's equations, and subsequent work showed how nonlinear gauge particle self-interactions must at low energies be described by Yang-Mills equations (for spin 1) or the nonlinear Einstein's equations (for spin 2) \cite{Boulware:1974sr}. His soft theorems have been revived recently in attempts to go beyond QFT to constrain interactions among particles. His approach also fits with string-theoretical ideas that predict massless particles of helicity smaller or equal than two but not higher.

 \item{\bf Electroweak unification}
 
On October 17, 1967 Weinberg submitted for publication what proved to be his most famous paper: `A Model of Leptons' \cite{Weinberg:1967tq}. In it he uses spontaneously broken continuous local (gauge) symmetries to describe the weak interactions. The idea came to him that the short range of the weak interactions might be due to the mechanism that Higgs, Brout, Englert, Kibble and others had identified \cite{HiggsMech} in which the spontaneous breaking of local symmetries gives the corresponding gauge particle a mass by absorbing the would-be Goldstone bosons of broken symmetries as their longitudinal components. 
 
The initial model focussed exclusively on the weak interactions of leptons (particles like electrons or neutrinos that do not take part in the strong interactions). He first identified the unbroken gauge group as $SU_\ssL(2)\times U_\ssY(1)$ in which the subscript $L$ means `left-handed', reflecting the experimental fact that weak interactions only act on the left-handed component of the spin-half particles involved. The $SU_\ssL(2)$ acts on a two dimensional vector space whose components are electrons and neutrinos. The subscript $Y$ in $U_\ssY(1)$ refers to `hypercharge'  which is a new gauge charge that in practice is a linear combination of electric charge and the diagonal generator, $T_3$, of $SU_\ssL(2)$. (This gauge group had also been arrived at by Sheldon Glashow -- without the spontaneous symmetry breaking -- several years earlier \cite{Glashow:1961tr}.)
 
He found that  $SU_\ssL(2)\times U_\ssY(1)$ is the biggest symmetry group satisfying these phenomenological  conditions\footnote{An extra $U(1)$, corresponding to lepton number was also possible but he rejected it as a gauge symmetry on the basis that there was no evidence for a gauge field for this symmetry.} that (as shown later) also survived quantization (\emph{i.e.}~was anomaly free). He introduced a doublet of scalar fields $H$ whose vacuum expectation value would break the symmetry down to an unbroken group $U_\EM(1)$ whose massless gauge particle would be the photon, with symmetry breaking pattern:
 \be
 SU_\ssL(2)\times U_\ssY(1) \rightarrow U_\EM(1) \,.
 \ee
This simplest choice implied the existence of three massive spin-1 particles, $W^\pm$ and $Z^0$, whose longitudinal polarizations arise by absorbing three of the original four scalar components of the complex doublet field $H$. The lone scalar field that remains was predicted as a new massive spinless particle, the Higgs field.  
 
This model -- and its later extension to include strongly interacting particles (what is now called the Standard Model of particle physics) is arguably one of the most successful theories in the history of science, achieving a consistent unified description of the previously known weak interactions (those due to $W$-boson exchange) and electromagnetism (QED) based on a gauge theory. It predicted the existence of new weak interactions -- `neutral currents', mediated by $Z^0$ exchange with a predicted strength relative to the $W$-mediated weak interactions. It predicted the masses of the massive gauge bosons  $W^\pm, Z^0$ and the Higgs particle in terms of their couplings, and predicted the strength of the Higgs couplings to all other particles in terms of the masses of these particles. All of these particles have since been found in experiments: neutral currents in 1973, the $W^\pm, Z^0$ gauge bosons in 1983 and the Higgs particle in 2012, all with the predicted masses and couplings \cite{Weinberg:1967tq}. The model also predicts a host of electroweak observables as functions of only a small handful parameters and these predictions have since been confirmed with spectacular precision.  

The Nobel committee awarded Weinberg the 1979 Nobel Prize in Physics for this work, shared with Salam (for a similar result the following year \cite{Salam:1968rm}) and Glashow (for the earlier work on $SU_\ssL(2) \times U_\ssY(1)$ models \cite{Glashow:1961tr}). Although this award came before the direct experimental production of $W$ and $Z$ bosons, by the end of the 1970s the indirect evidence for the electroweak model had already become compelling. But direct confirmation eventually did come, with experimental evidence for the last hitherto undiscovered particle -- the Higgs boson -- detected at CERN in 2012 (with all current evidence indicating it to have precisely the properties predicted for the Higgs particle in the work of Weinberg and Salam.\footnote{Higgs and Englert would later share the prize for the discovery of the Higgs \emph{mechanism} and the observation that it often comes associated with spinless particles like the Higgs boson of the Weinberg-Salam model.} 
 
 \end{itemize}
 
\subsection{1970s}

Nothing much was really done with the model of leptons for several years immediately after publication. This changed in the early 1970s when it was quickly extended to include strongly interacting particles, leading to the Standard Model (SM) we know today. The delay in exploitation had its roots in one of the model's key assumptions: that the model should be renormalizable. Renormalizability means that all ultraviolet (UV) divergences can be absorbed into the model's parameters and this was understood to be necessary for being able to make precise predictions (and so was a prerequisite for making sensible comparisons with experiments). Although spontaneously broken gauge theories were suspected to be renormalizable, the proof of this did not come until Gerard 't Hooft proved it\footnote{It is interesting that Weinberg was \emph{not} the person to prove this. In retrospect he was thwarted by his use of unitary gauge -- a convenient choice for studying the model's phenomenology but a perverse choice for studying its UV properties -- to study the implications of the model. He failed to appreciate that by using the path integral formalism more convenient gauge choices were possible \cite{Weinberg:2004kv}.} in 1971 \cite{tHooft:1972tcz}. Once renormalizability was established the credibility of the model increased substantially, stimulating an explosion of activity that ended with a picture of modern particle physics that largely survives unscathed up to the present day. An inspiring video description of this achievement told by Weinberg himself can be found in \cite{colloquium} (see also \cite{Weinberg:2004kv, Weinberg:2018apv}).

Weinberg played a central, but no longer quite so solitary, role in this explosion. On the Standard Model side, he made seminal contributions to the modern emergence of Quantum Chromodynamics (QCD) as the gauge theory of the strong `colour' interactions amongst quarks, and to its ultimate combination with the electroweak model to produce the Standard Model, based on the gauge group $SU_c(3) \times SU_\ssL(2) \times U_\ssY(1)$. He also played a pivotal role in the search for the higher-energy completions of the Standard Model -- what has come to be known as \emph{Beyond the Standard Model} (or BSM) physics. He was among the first to study the implications of both Standard and BSM physics for cosmic evolution in the very early universe, and by so doing helped make the study of cosmology a mainstream particle-physics passtime. A further major intellectual thread that winds through his research at this time is the extension and maturing of his formulation of Effective Field Theories (EFTs) and the key role they play in making reliable predictions using nonrenormalizable theories (like General Relativity). By so doing they also explain why renormalizable theories are so prominent amongst our successful descriptions of nature.

\begin{itemize}

\item{\bf Emergence of the Standard Model}

Weinberg's main contributions to the maturing of the Standard Model from just a model of leptons were to do with formulating how the strong interactions fit into the picture.   

\begin{itemize}

\item {\bf  QCD: massless gluons} 

During the 1960s evidence accumulated that strongly interacting particles (protons, neutrons, pions and so on -- collectively: hadrons) could be described as bound states of more fundamental objects (quarks) carrying a new quantum number called `colour' corresponding to a symmetry group $SU_c(3)$. Evidence also accumulated that although the interactions of these quarks were strong enough at low energies to bind them together into hadrons, they also seemed to become relatively weak at the highest energies produced in the colliders of the time. These were important clues about the nature of the strong interactions, but it was initially hard to use them. 

Although it was understood that the apparent strength of an interaction could change at different energies, it was hard to find examples for which the higher energy interactions are weaker than those at lower energies (what is called `asymptotic freedom'). The picture became clearer after Gross, Politzer and Wilczek \cite{AsymptoticFreedom} showed that a nonabelian gauge theory (the kind of gauge theory required if it is to couple to colour charges) could be asymptotically free. This led to the widespread acceptance that the strong interactions might also be described by a gauge theory, based on the symmetry group $SU_c(3)$.

There were two important conceptual puzzles with this picture. One of these was the short range of the strong interactions, which seemed to suggest that the corresponding gauge bosons -- for the strong interactions these are called `gluons' -- must be massive, similar to the way the $W$ and $Z$ boson masses give the weak interactions a finite range (and the masslessness of the photon corresponds to the infinite range of the Coulomb force). The second puzzle was why the consituent quarks were never seen outside of a hadron. We know of many kinds of bound states in Nature (\emph{e.g.}~atoms) but normally we also find their free constituents (electrons and nuclei) in experiments as well. Free quarks were never found, however, even after hadrons were collided with much more energy than their typical binding energy. The absence of free quarks was (and still is) believed to be because nonabelian gauge theories \emph{confine}: they make pulling a quark out of a hadron energetically more expensive than simply pulling an entire hadron/anti-hadron pair from the vacuum. 

In 1973 Weinberg proposed what is now the mainstream view: that the gluons could actually be massless and this need not be inconsistent with the strong force having a finite range \cite{Weinberg:1973un}. Finite range is consistent with massless gluons because gluons are also confined (because they also carry the colour charge), and the range of the strong force is instead set by the mass of the lightest colour-neutral hadron (in practice the pion) that can be exchanged.  

\pagebreak
\item{\bf  Jets and IR safe observables}

Characteristically, Weinberg's foray into the strong interactions also had a practical calculational aspect. Although having couplings be weak at high energies in principle allows perturbative methods to be used to compute the results of scattering processes, these processes are only simple for quarks and gluons. But these are not what experiments actually observe. Even worse: having gluons be massless and also carry the colour charge raised another problem: infrared (IR) divergences. The arguments that IR divergences in electromagnetism cancel once photon emission from final states is taken into account actually turn out to fail if there are charged particles that are also massless. Since gluons are both massless and carry colour charge infrared divergences involving the strong interactions are not cancelled simply by summing over soft gluon emission from final-state particles. Instead these divergences only cancel once one sums over soft gluon emission from \emph{both} the initial- and final-state particles (a result known as the KLN theorem, for Kinoshita, Lee and Nauenberg \cite{Kinoshita, LeeNauenberg}). 

Weinberg, together with George Sterman \cite{Sterman:1977wj}, showed how practical calculations relevant to experiments can be performed for collisions involving strongly interacting quarks and gluons: the trick was to compute the rate for producing at least one energetic (`hard') particle plus a sum over an indeterminate number of soft gluons and quarks travelling in roughly the same direction. Such bundles of particles -- called `jets' -- are both easy to compute perturbatively (at high enough energies) in terms of quarks and gluons and easy to measure (one simply sums over all particles heading in a particular direction above a certain energy threshold). They are also IR safe because of the sum over all quark and gluon fellow travellers both within the final-state jet or jets and within the particles being collided (weighted by their initial wave-function). This observation provided a practical way to infer when hard scattering of quarks and gluons occurs in particle colliders, making jets of hadrons the variables of choice when interpreting experiments. 

\item{\bf  Finite temperature quantum field theory}

Following the observation of Kirzhnitz and Linde \cite{Kirzhnits:1972ut} that spontaneously broken symmetries can be restored in the high temperatures of the early universe for non-gauge theories, Weinberg \cite{Weinberg:1974hy}  --  simultaneously with Dolan and Jackiw \cite{Dolan:1973qd} -- proved that the same also happens for gauge symmetries. This predicted an interesting phase-transition history for the universe, in which symmetries are restored in the extreme temperatures of the very early universe and so contain a very different spectrum of particles than are currently present. Phase transitions occur as the universe expands and cools, leaving in the end the broken symmetry seen around us at present-day low temperature, allowing the spectrum of particles available in the very early universe to be completely different from those we see now.

\end{itemize}

\item {\bf Beyond the Standard Model (BSM)}

Alternatives to the Standard Model were being sought before the ink dried on the papers that proposed it. At first these searches were not driven by disagreements between the Standard Model and observations. They were rather more of a fishing expedition: once it became clear that the strong and electroweak interactions were described by renormalizable gauge theories, and it was clear that more symmetries were possible at higher energies, it was inevitable to explore different types of gauge theories to see what Nature's options might have been.

\pagebreak
\begin{itemize} 

\item{\bf  Gauge coupling unification}

One of the initial BSM directions proposed that the additional gauge symmetries that are possible at high energy might include interactions that unify the strong with the electroweak interactions -- what are now known as Grand Unified Theories (GUTs). These interactions appear as two separate pieces of the Standard Model but they might only appear to be different due to spontaneous symmetry breaking at much higher energies, in much the same way (within the Standard Model) that the weak and electromagnetic interactions seem very different at low energies but look very similar at high energies. 

Early work along these lines by Pati and Salam \cite{Pati:1974yy}, Georgi and Glashow \cite{Georgi:1974sy} and others proposed a simple (in the technical sense) symmetry group, like $SU(5)$, that breaks at high energies to the symmetries of the Standard Model. Because the new unifying interactions often cause transitions between quarks and leptons, they can generically provide the proton with a means to decay ({\it e.g.}~through a process like $p \to e^+ \pi^0$). The very strong evidence for the absence of proton decays implied the scale of this symmetry breaking must be extremely large: at least $10^{13}$ times higher than the Standard Model's scale of electroweak symmetry breaking.

Weinberg -- with Howard Georgi and Helen Quinn -- provided a second, independent, indirect argument for the existence of such a high scale for GUT symmetry breaking \cite{Georgi:1974yf}. They observed that unification predicts the strengths of the electromagnetic, weak and strong interactions to be equal at the symmetry breaking scale, despite their being quite different in size at the lower energies where we can measure them. This need not be a problem because coupling strengths do evolve with energy, but this running is calculable subject to assumptions about the kinds of particles that exist with masses between the high GUT scale and ordinary energies. 

Weinberg and friends computed this running using the simplest assumption: use only the known Standard Model particles. They found two remarkable things: first that it is indeed possible for the measured couplings to unify at high enough energies, in the sense that there exists an energy scale where they all become the same size -- within measurement errors. (This need not happen in general: running couplings can be graphed as curves as a function of energy and although two weakly curved lines with different slopes eventually cross, there is no guarantee that a third line also does so at the same place.) Second, because the running is only logarithmic in energy the unification scale to which one is led is very high, turning out to be satisfyingly consistent with the values required for acceptable suppression of proton decay. 

By the 1990s the precision of measurements of coupling strengths had improved to the point that it was no longer true that a scale exists for which the strong, weak and electromagnetic couplings could be equal within the errors, if evolved using only Standard Model particles. It was soon realized that supplementing the known Standard Model particles with the new particles predicted by supersymmetric theories (proposed for completely different reasons to do with the Hierarchy Problem -- more about which below) restores the ability to have the observed couplings unify, doing so at a slightly higher scale. The possibility that this might not just be a coincidence remains a tantalizing hint for the existence of physics beyond the Standard Model.
 
Within the context of cosmology, having unified symmetries at high energies also means the very early universe would be symmetric under the GUT symmetry, adding to the phase transitions for which observational consequences in the later universe can be sought. A particularly important class of such consequences led to theories of \emph{cosmic inflation} \cite{Guth:1980zm, Linde:1981mu, Albrecht:1982wi}), in which many of the puzzles associated with the initial conditions required by successful Big Bang cosmology are explained by the accelerated expansion that can occur near a phase transition in the much earlier universe\footnote{Weinberg had a major influence on the original proposal by Alan Guth of the inflationary universe: Guth learned cosmology reading both of Weinberg's books (his textbook on Gravitation and `The First Three Minutes') and was motivated to address cosmology questions for GUTs from the particle physics perspective by attending a seminar of Weinberg on baryogenesis\cite{guth}.  }. Tantalizingly, success of these cosmologies often turns out to require that this expansion should occur for temperatures as high as the GUT symmetry breaking scale. 

\item{\bf  The Hierarchy Problem}

Grand Unification led to a picture within which the Standard Model emerges at low energies from something more fundamental at much higher energies due to a pattern of spontaneous symmetry breaking. With his graduate student, Eidad Gildener \cite{Gildener:1976ih} Weinberg identified a fundamental conceptual problem --  what came to be known as the \emph{electroweak hierarchy problem} -- with having this symmetry breaking happen due to a vacuum expectation value for an order parameter that is a weakly coupled elementary scalar field (as happens with the Higgs field in the Standard Model). 

Any such theory would require elementary scalar fields with vastly different masses: some collection of very heavy fields to break the high-energy gauge group and the low-energy Standard Model Higgs field to break the electroweak symmetry. The problem is that theories with elementary scalar fields tend not to produce masses for those fields that are enormously different in size. With the exception of Goldstone bosons scalar fields tend to be very sensitive to quantum fluctuations and this makes it rare to find large mass hierarchies amongst scalars. 

A similar thing is found in \emph{e.g.}~condensed matter physics, where scalar order parameters are typically only gapless very near to critical points. For instance the phase diagram of water has a critical point at a temperature near 374° C, a scale characteristic of the microscopic interactions between water molecules. Although the existence of critical points at nonzero temperature is not rare, it is much more unusual to find quantum critical points that are critical very near to zero temperature, and when these are found there is usually a mechanism at work that suppresses the critical temperature relative to the scales of the underlying interactions. The Standard Model within a GUT is like a quantum critical point in this way; the scale of electroweak symmetry breaking is so small compared with the GUT scale that something needs explaining.

\item{\bf  Technicolour}

Having identified the hierarchy problem for the Higgs particle, Weinberg \cite{Weinberg:1975gm} and independently Susskind \cite{Susskind:1978ms} proposed a way out: the order parameter that breaks the electroweak symmetry (and the Higgs boson itself) might not be an elementary scalar. They proposed the existence of a new gauge interaction -- called technicolor -- that binds constituent fermions together to form the Higgs in much the same way that QCD binds quarks together to make pions, or similar to the Cooper pair breaking of the electromagnetic $U_\EM(1)$ symmetry in superconductors.

\item{\bf  Axions}

The observation that the SM might be the low-energy limit of something more fundamental at much higher energies stimulated much searching for SM properties that would not normally be expected to be present at very low energies. One such a property is the apparent absence of CP violation in the strong interactions, a quantity controlled in the SM by a dimensionless parameter $\Theta$ that must be smaller than of order $10^{-10}$ in order to agree with experiments (such as the observed absence of a neutron electric dipole moment). The small size required for $\Theta$ became known as the \emph{strong CP problem}.

Helen Quinn and Roberto Peccei proposed \cite{Peccei:1977hh} an elegant explanation for why $\Theta$ could be small in a natural way, involving the introduction of a new spontaneously broken global symmetry $U_\PQ(1)$. Weinberg \cite{Weinberg:1977ma} -- and independently Frank Wilczek \cite{Wilczek:1977pj} -- realized that this symmetry breaking pattern requires the existence of a very light new particle: the Goldstone boson for this symmetry breaking (now called the \emph{axion}), transforming under it as $a\rightarrow a + $ constant. 

The axion is what tells the low-energy world about the existence of the spontaneously broken new symmetry, and so plays an instrumental role in the symmetry's observable consequences. Many dedicated experimental searches have been designed to seek evidence for (or rule out) its existence. To this day this QCD axion remains one of the best-motivated candidates to explain the origins of the unknown Dark Matter that astrophysical and cosmological observations indicate is a major component of the current energy density of the universe. The word `axion' has also come to be more broadly applied to any axion-like particle (or ALP); that is, any sort of Goldstone boson appearing at low energies, even if its couplings are not the precise ones that solve the strong CP problem. This broader class of axions is widely studied both as a candidate for Dark Matter and also because they commonly arise in well-motivated theories like string theory (more about which below).


\item{\bf BSM Cosmology}

Weinberg was early to appreciate the utility of cosmology for particle physics, and his example was influential in bringing cosmological arguments into the mainstream. 

{\it Neutrino cosmology: }
A powerful technique he helped pioneer was the use of cosmological observations to constrain the properties of particle physics models. 
Together with his good friend Benjamin Lee, Weinberg explored the cosmological implications of heavy neutrinos \cite{Lee:1977ua}, finding a constraint on neutrino mass that must be satisfied if its cosmic abundance is not to exceed the current energy density of the universe. The Lee-Weinberg bound forms the basis for discussions of other candidates for cold dark matter, such as WIMPS (weakly interacting massive particles), because it provides a way to relate a particle's mass and coupling to its relic density in the late universe. If initially in thermal equilibrium the abundance of the dark matter particle eventually freezes-out as the universe expands and cools (since these reduce their annihilation rate ), at a temperature  that can be computed as a function of the particle's mass and couplings. The predicted relic density can be a boon (if it gives the observed abundance of Dark Matter) or a bane (if it is so high that it overwhelms other sources of energy and over-closes the universe).

{\it Baryogenesis:}
The appearance of baryon-number violating interactions at very high energies within Grand Unified Theories suggests that these types of interactions could have occurred more frequently in the very early universe when temperatures were much higher. This in turn opens up an opportunity to understand the baryon asymmetry of the universe (why it currently contains mostly baryons and hardly any antibaryons), as several groups independently realized 
\cite{Dimopoulos:1978kv, Toussaint:1978br, Weinberg:1979bt}. Weinberg's contribution \cite{Weinberg:1979bt} was very influential, showing in particular how the baryon asymmetry (normalized to the CMB's photon entropy, since both scale with universal expansion in the same way) could be computed in terms of microscopic parameters of the GUT model of interest. 

\pagebreak
{\it  Ultimate early universe (Hagedorn) temperature:}
One of Weinberg's earlier papers on cosmology was unusually speculative \cite{Huang:1970iq} and written in 1970 (well before the establishment of the Standard Model). Even though not attracted to work on popular alternatives to QFT -- like $S$-matrix theory and `bootstrap' methods -- that were being widely explored at the time, he nevertheless -- with Kerson Huang -- wrote a very influential paper about the ultimate temperature (known as the Hagedorn temperature) that arises in theories (like string theory) whose density of states grows exponentially at high energies. In this paper they compute the critical temperature and study further potential implications of an ultimate temperature for cosmology. 

\item{\bf  Asymptotic Safety}

Building on the great success of asymptotic freedom in QCD, Weinberg conjectured the existence of a broader class of theories -- which he called \emph{asymptotically safe} -- that might be similarly predictive. The idea was that rather than having a theory's couplings flow at high energies to a fixed point, it might happen that couplings end up flowing towards a finite-dimensional (say, $N$-dimensional) surface within coupling-constant space. In this case the theory remains predictive because of the UV information that the couplings can be specified at very high energies by giving only $N$ pieces of information.  Asymptotic freedom would then be regarded as a special case where a single coupling goes to a specific value ($N=0$) at high energies. 

Gravity is what provided his motivation for thinking along these lines. General Relativity (GR) is well-known to be a non-renormalizable theory, which means that the UV divergences that arise when extracting its physical predictions cannot all be absorbed into the theory's coupling constants (like Newton's gravitational constant $G_\ssN$ for GR). At face value this problem can be resolved by adding new coupling constants, such as would arise if the Einstein-Hilbert action of GR 
\be
  S_{\scriptscriptstyle GR} = - \frac{1}{16 \pi G_\ssN} \int \exd^4x \; \sqrt{-g} \; R  \,,
\ee
were replaced by a curvature expansion that includes higher powers of $R$ (and all other measures of curvature),
\be
   \widehat S = - \frac{1}{16 \pi G} \int \exd^4x \; \sqrt{-g} \; \Bigl[ R + c_1 R^2 + c_2 R_{\mu\nu} R^{\mu\nu} + \cdots \Bigr] \,,
\ee
where $R_{\mu\nu}$ and $R$ are measures of the spacetime metric's curvature. Once this is done UV divergences \emph{can} all be absorbed into the new couplings $c_i$. But the problem then is predictivity: in principle there are an infinite number of parameters to be determined when making predictions, particularly for high energies and strong curvatures since then keeping only the first term in a curvature expansion should not be a good approximation.

But predictability could be restored \emph{even at high energies} if gravity turned out to be asymptotically safe. Then the couplings $c_i$ would all asymptote towards an $N$-dimensional surface in the space of couplings and so would not all be independent of one another, allowing predictions to be made once $N$ observables are used to pin down the value of the $N$ unknown couplings. 

Weinberg explored this idea when thinking about what effective field theories can say about the problem of gravity's nonrenormalizability (more about which below), and did not immediately write a paper fleshing out these ideas -- instead describing them in his 1976 lectures on critical phenomena \cite{safetyfirst} and following up in his 1979 contribution to the book honoring the centennial of Einstein's birth \cite{safetysecond}. The proposal was eventually picked up by others (see for instance, \cite{Kawai:1993mb, Dupuis:2020fhh}, spawning much activity -- both trying to accumulate evidence for asymptotic safety and seeking phenomenological consequences (including a return to the subject by Weinberg himself \cite{Weinberg:2009wa}). 

It remains controversial whether gravity in 4 spacetime dimensions is actually asymptotically safe (our main calculational tools are restricted to the low-energy and weak-curvature regime), but if it were then standard QFT techniques could suffice to describe gravity in the strongly quantum domain (and thereby provide the simplest approach to making sensible predictions for it at high energies).  

\end{itemize}

\item{\bf  Effective field theories}

It was during the 1970s that Weinberg's vision of the power and ubiquity of effective field theories (EFTs) started to be fleshed out in full generality \cite{Weinberg:1978kz, Weinberg:2021exr}. This was in part driven by the discovery of the Standard Model and the desire to extract its predictions across the wide range of mass scales to which it applies. Although initially developed to understand pion interactions at low energies compared to the GeV scales typical of hadron masses, EFT methods work equally well when describing the effects of virtual $W$ and $Z$ bosons at energies below their masses and when used in this way provide a firmer theoretical foundation for the earlier phenomenological Fermi theory (and its later parity-violating generalization) of the weak interactions. But  appreciating their generality -- they apply in \emph{any} situation with a hierarchy of scales -- reveals many other novel places where they are also useful, such as identifying the low-energy influence of virtual muons and/or precision QED radiative corrections within atoms whose binding energies are well below the electron mass \cite{Caswell:1985ui}.

It soon also became clear that the efficiency of EFTs in dealing with hierarchies of scale was also useful if the Standard Model itself is the low-energy limit of some more fundamental theory (such as a Grand Unified Theory) at much higher energies. The key observation is that the low-energy influence of very massive particles with mass $\Lambda$ can be universally captured through a series of local interactions in an effective lagrangian involving only the light degrees of freedom. For instance the virtual effects of a heavy particle on a light scalar field $\phi$ are well-described at energies $E \ll \Lambda$ by an effective lagrangian of the form
\begin{equation}
{\mathcal L}_{\text{eff}}=\underbrace{\underbrace{-\partial^\mu\phi \, \partial_\mu \phi - m^2\phi^2-g\phi^3-\lambda\phi^4}_{\text{Renormalizable}} - \frac{\alpha}{\Lambda}\phi^5 - \frac{\beta}{\Lambda^2}\phi^6+\ldots}_{\text{Non-Renormalizable}}\,,
\end{equation}
where the ellipses involve all possible local operators involving the light field. All interactions are local to any fixed order in $1/\Lambda$, ultimately because of the interplay between energy-momentum conservation and the uncertainty principle: any virtual high-energy particle can only appear at low-energies because of the violation of energy/momentum conservation that the uncertainty principle allows. But this implies the effect takes place only over very short times/distances.\footnote{This description is appropriate for old-fashion Raleigh-Schr\"odinger perturbation theory where particles are always `on shell'. A slightly different (but equivalent) rationale involving off-shell dispersion relations can be given within relativistic Feynman-Schwinger perturbation theory.} 

It is noteworthy that all but the first few (renormalizable) interactions are suppressed by powers of $1/\Lambda$ on dimensional grounds, showing that the renormalizable terms contain the only low-energy effects of virtual heavy particles that are not suppressed by a power of $\Lambda$. This explains why renormalizability plays such an important role in our successful descriptions of nature (such as for QED, QCD or the entire Standard Model itself). Renormalizable theories are what emerge at low energies as the leading description since they express all of the low-energy predictions that are not suppressed by a power of $E/\Lambda$. If the particles and gauge symmetries in the low-energy theory with energies $E \ll \Lambda$ are those found in the Standard Model then we should expect their interactions to be described at low energies by the most general renormalizable theory consistent with them. That is to say: by the Standard Model itself. Although the particles and symmetries of the Standard Model have a beauty that only a mother could love, the \emph{inevitability} of the Standard Model at low energies (given the particle/symmetry content) provides a truly beautiful explanation for \emph{why} the Standard Model works so well.

This same picture also provides a way to make sense of non-renormalizable theories (like General Relativity) in a quantum context. This might be surprising because GR involves only massless particles and so one might think that there is no hierarchy of scales for EFTs to exploit. However, all nonrenormalizable theories have coupling constants that have dimension of inverse powers of mass (in fundamental units for which $\hbar = c = 1$). For instance, Newton's gravitational coupling is $G_\ssN = 1/M_p^2$ where $M_p \sim 10^{19}$ GeV is the Planck mass, much like the Fermi constant $G_\ssF$ of low-energy weak interactions below the $W$ mass is proportoinal to $\alpha/M_\ssW^2$ (where $\alpha$ is the electromagnetic fine-structure constant). Quantum calculations with these theories turn out in practice to be an expansion in powers of energy divided by this mass scale. With this point of view quantum calculations within gravity are under control and predictable provided they are done only at low energies compared with scales like $M_p$ (which is certainly valid for all processes currently amenable to experiments).

The point of EFT methods is that it is much more efficient to perform a low-energy expansion directly in a system's Hamiltonian (which then can be applied to \emph{any} low-energy observable) than it is to first compute each observable and then expand the full result. This improved efficiency of calculation often allows calculations to be performed that were hitherto impractical to do. As a consequence subsequent decades have seen the relentless spread of Weinberg's EFT tools throughout physics. 

\begin{itemize}
\item{\bf First steps in SMEFT}

Weinberg himself provided the first tantalizing application of these tools to BSM physics. If all non-standard particles in a particular theory are heavy compared to the $W$-boson mass, say, then their effects at accessible energies must be expressible in terms of a collection of non-renormalizable interactions built using only SM fields (an effective field theory now known as SMEFT). Different underlying models can only differ at low energies in the values they predict for the various effective couplings. Studying the observable consequences of these effective interactions allows a relatively model-independent exploration of those theories that share the SM particle content at low energies.

Weinberg began this kind of analysis by studying processes that violate conservation of Baryon number ($B$) or Lepton number ($L$). These conservation laws automatically fall out as predictions in the Standard Model despite not being assumed as one of the assumptions in the model's definition. Consequently \emph{any} microscopic theory (like Grand Unified Theories) having only SM states at low energies must accidentally conserve $B$ and $L$ so long as $1/\Lambda$ effects can be neglected, since in this limit only the renormalizable SM interactions can contribute. This shows how $B$ and $L$ conservation emerge as accidental low-energy consequences of renormalizability given the right low-energy particle content and gauge symmetries. 

Starting from this observation, Weinberg classified all of the possible $B$ and $L$ violating interactions that are possible at lowest order in $1/\Lambda$ using only SM particles and gauge symmetries (but no longer requiring renormalizability, as appropriate for SMEFT) \cite{Weinberg:1979sa}. Writing 
\begin{equation}
\cL=\cL_{\SM}+ \cL_{5}+ \cL_{6}+ \cdots\,,
\end{equation}
where $\cL_d$ for $d \geq 5$ contains all possible interactions with dimension (mass)${}^d$, implies that all couplings in $\cL_d$ are proportional to $\Lambda^{4-d}$. The larger $d$ is the more suppressed at low energies the interaction should be expected to be. 

Only a unique interaction is allowed for $d=5$ ($B$ and $L$ violating or not), with
\begin{equation}
\cL_5=  \dfrac{\lambda_{ij}}{\Lambda} \; HH ( L_i   L_j ) + \hbox{c.c.} \,, 
\end{equation}
where $i,j$ are generation indices (distinguishing for example electrons, muons and tau leptons) and $\lambda_{ij}$ is a symmetric complex matrix of dimensionless couplings. $H$ denotes the Higgs doublet and $L_i$ denotes the three generations of left-handed lepton doublets. Gauge and Lorentz-spinor indices are suppressed but are contracted in the only way allowed by $SU_\ssL(2) \times U_\ssY(1)$ gauge invariance and local Lorentz invariance. 

This interaction is what would be expected to capture the leading deviations from SM predictions if $\Lambda$ is very large and its presence would manifest itself experimentally mainly by giving neutrinos masses that are order $v^2/\Lambda$ in size, where $v \simeq 246$ GeV is the Higgs expectation value. These masses would be expected to be very small if $\Lambda$ were large, being of order $10^{-3}$ eV if the $\lambda_{ij}$ couplings are order unity and $\Lambda \sim 10^{16}$ GeV is close to the GUT scale. Remarkably, we now know that among the very few experiments contradicting the SM are those that indicate that neutrinos have masses, with experiments since the 1990s providing ever-more-convincing evidence for masses in the ballpark of $10^{-3}$ eV. 

Weinberg also classified the most general $B$ and $L$ violating interactions that arise in $\cL_6$, finding a relatively small list with the schematic form
\begin{equation}
\cL_6= \dfrac{\beta_{ijkl}}{\Lambda^2}  \Bigl( qqq \ell \Bigr)
\end{equation}
where $q$ denotes either a left-handed or right-handed quark field and $\ell$ denotes either a left-handed or right-handed lepton.  There turn out to be six independent ways these types of fields can be combined into an $SU_c(3) \times SU_\ssL(2) \times U_\ssY(1)$ invariant combination. All of these interactions satisfy the selection rules $\Delta B = \pm 1$ and $\Delta B = \Delta L$, and so these must also be selection rules for \emph{any} GUT that has only SM particles at low energies. For any such a model protons and neutrons can decay but they must always decay into antileptons (so decays like $p \to \pi^0 \ell^+$ or $n \to \pi^- \ell^+$ are allowed but decays like $p \to \pi^0 \pi^+$ or $n \to \pi^+ \ell^-$ or $n - \bar{n}$ oscillations are forbidden at order $1/\Lambda^2$ in the amplitude).

\end{itemize}

Looking back, these papers on EFT methods and their application to SMEFT are seen to be seminal contributions that still have a very modern feel. Partly this is because these papers helped to define what we think of as a modern analysis. In the intervening years since the late 70s we have come to learn that collider experiments continue to support the validity of the Standard Model, out to energies of order $10^4$ GeV. The absence of evidence for new exotic particles in these experiments (and the discovery of the Higgs boson in 2012) have made SMEFT a standard tool to analyze BSM physics.

\end{itemize}

\subsection{1980s}

After the explosive development of the Standard Model in the 1970s the 1980s were more a time of consolidation, with more of a theoretical focus on searches for BSM physics, often using the framing of the electroweak hierarchy problem as a central clue. 

It was also during the 1980s that the importance of gravity to these questions came into focus, making a working knowledge of General Relativity a mandatory part of a particle physicist's training. From a phenomenological point of view the need for gravity emerged as it became clear that BSM physics often drew on physics at very high energies -- not too far below the Planck scale $M_p$ -- for which gravitational effects need no longer be negligible.

But it was also around this time that string theory -- and the extra-dimensional Kaluza Klein theories to which it leads -- became widely appreciated as a viable candidate for a quantum theory of gravity even at arbitrarily high energies (what is known as a `UV completion' of General Relativity). String theory differs from ordinary particle theories by having an infinite tower of states (in the weak-coupling limit these are the oscillation modes of the strings) spaced by a characteristic mass scale $M_s$, which was also plausibly only a little below $M_p$ for weakly coupled strings. Having such a tower precludes describing physics above $M_s$ in terms of a conventional effective field theory, and opens up the possibility for much better UV behaviour. Strings always contain a massless spin-two particle mode that plays the role of the graviton, and interacts with other light particles by exchanging all possible virtual heavy string modes. This exchange allows Newton's constant $G_\ssN$ to be computed as a function of $M_s$ and the string coupling, $g_s$, in much the same way that the Fermi coupling $G_\ssF$ can be computed in the Standard Model in terms of the fine-structure constant $\alpha$ and the $W$-boson mass $M_\ssW$. 

Although Weinberg dabbled in, and kept abreast of, string theory he did not concentrate his work there, judging he would likely have more impact in other directions. He did influence later developments in string theory and extra dimensions and strongly supported research in this area, which he considered the most promising approach towards quantum gravity (including his own asymptotic safety proposal). In his own words:
\begin{quote}
{\it If I have to bet my life I would bet on string theory rather than asymptotic safety\cite{armas}.}
\end{quote}

Some highlights of his research in this decade now follow. 

\begin{itemize}

\item{\bf The Weinberg-Witten Theorem}

In 1980, while still at Harvard, Weinberg found a rare collaborator whose name came later in the alphabet than his own (in particle physics papers authors are usually listed alphabetically): a young Edward Witten. They extended Weinberg's own earlier work on gauge particles with higher spin; proving a very general theorem constraining massless particles \cite{Weinberg:1980kq}. The theorem states that  in all theories with a Lorentz covariant energy-momentum tensor, massless particles of helicity $j>1$ are forbidden, a result that in particular rules out the graviton being a composite particle. 

This theorem attracted renewed attention some 15 years later in the context of AdS/CFT dualities; theories for which the graviton emerges as a state within an ordinary Yang-Mills gauge theory. This is possible in this case because in these theories the graviton in $D$ spacetime dimensions arises as a state within the gauge theory localized at the $(D-1)$-dimensional boundary of spacetime, with the different dimensions threading a loophole of the Weinberg-Witten theorem.

\item{\bf Supersymmetry}

Weinberg did not work on developing supersymmetric theories during the 1970s as much as did many of his peers. His interest was eventually engaged by its possible utility in helping deal with the electroweak hierarchy problem. Supersymmetry relates the properties of spinless particles to those with spin-$\frac12$ and so can tie the mass of elementary scalars to those of fermions (whose masses can be kept small using standard symmetry arguments). The hope was to use this type of observation to find theories for which the Standard Model Higgs could be kept much lighter than any much higher fundamental scales. 

His early work in this area sets out a very general approach for pursuing the phenomenological implications of models with rigid\footnote{For rigid symmetries transformation parameters are independent of spacetime position, as opposed to gauge symmetries where the parameters can vary in space and time.} supersymmetry \cite{Weinberg:1981wj}, finding that there are many obstacles to making such models work when supersymmetry is spontaneously broken at energies just beyond our experimental reach. These studies suggested the scale of supersymmetry breaking would likely be at much higher energies, within a sector coupled much more weakly to ordinary particles (what came to be known as a \emph{hidden sector}) \cite{Polchinski:1982an}. 

Weinberg wrote several influential articles on the possible observational consequences of supergravity. He immediately realised that there is a cosmological limit on the mass of the gravitino\footnote{The gravitino is the spin-$\frac32$ gauge particle for local supersymmetry transformations.} \cite{Weinberg:1982zq}. If unstable, the gravitino must be heavy enough to decay before the epoch of Big Bang nucleosynthesis (BBN). But the more massive the gravitino the higher in energy must be the scale of supersymmetry breaking, pushing it into the hidden-sector regime. Once supersymmetry-breaking scales are pushed high enough gravity can no longer be reliably neglected. This is both a blessing and a curse: it is a blessing because gravitational interactions are inevitable, and so gravitational strength interactions with the hidden sector always exist and do not require the invention of new types of weak interactions. It is a curse because the formalism of supergravity is more complicated than is the formalism for rigid supersymmetry. 

Not being one to fear useful formalism, Weinberg and collaborators founded the field of gravitational mediation of supersymmetry breaking by providing a general analysis of supergravity models with high scale supersymmetry breaking mediated by a gravitationally coupled sector \cite{Hall:1983iz}. This allows a large scale of supersymmetry breaking, $M_s$, while still only giving $M_s^2/M_p \sim $TeV scale masses to the superpartners of the Standard Model particles. Such models address the hierarchy problem because supersymmetry protects the Standard Model Higgs from becoming too massive due to the approximate cancellation of high-energy fermions and bosons in its mass.

In later years these gravitational hidden-sector models saw a revival, but this time from theorists working down in energy starting from string theory (as opposed to working up in energy, starting from electroweak phenomenology), as supersymmetric compactifications of string theory were obtained and often found to contain such hidden sectors, containing (for instance) the fields (moduli) determining the size and shape of the extra dimensions.  

\item{\bf Higher dimensions}

In the early 1980s the emergence of string theory as the first known candidate for a UV completion of General Relativity (but only in $D \geq 10$ dimensions, at least for weak coupling) combined with the low-energy realization a few years earlier that supersymmetry could play a role explaining why the Standard Model could be much lighter than gravity-related scales like $M_p$, led to an explosion of research into the dynamics of higher-dimensional supergravity theories. 

It was known (since the days of Kaluza and Klein) that such theories could be consistent with observations showing us to live only in 4 spacetime dimensions provided any `extra' dimensions were sufficiently small. An important question for these models was therefore to compute the size and shape of these extra dimensions, to see why they should be so small. One way to do this was to postulate the existence of additional fields, whose quantized flux within an extra-dimensional geometry could counter gravity's tendency to collapse extra dimensions to ever-smaller volumes \cite{Freund:1980xh}. 

Weinberg, with Philip Candelas provided among the first calculations where it is instead quantum effects -- Casimir energies within the extra dimensions -- that compensate the curvature of the extra dimensions \cite{Candelas:1983ae}. Most interestingly this is also among the first examples where gravitating quantum effects were computed \emph{in a controlled way}, as justified by EFT arguments, without simply assuming the metric must be a classical geometry on which other quantum fields move. In principle the graviton Casimir effect also constributes to the stabilization of extra dimensions, though Candelas and Weinberg argued that this can be neglected in the limit where there is a large number of other fields. This line of research ultimately led Candelas, together with Gary Horowitz, Andrew Strominger and Edward Witten \cite{Candelas:1985en}, to find the celebrated Calabi-Yau solutions that remain the leading approach towards studying supersymmetric 4D string vacua 40 years later. 

\item{\bf The cosmological constant problem}

In 1988 Weinberg gave the Morris Loeb Lectures in Physics at  Harvard University, choosing as his topic the {\it Cosmological Constant Problem}. This is another type of hierarchy problem, which asks why the observed energy density of the vacuum (at the time of Weinberg's writing the vacuum energy was consistent with zero, it is now known to be around $10^{-120}$ in Planck units) should be so small compared with all of the other fundamental scales in physics. Very small vacuum energies are unusual at low-energies in theories with much heavier particles in much the same way that very small scalar masses also can be. Unlike for scalar masses all of the known symmetries that might help seem to have other implications that rule them out, so no known theories convincingly resolve this problem.\footnote{At the time Weinberg gave his lectures the most actively pursued approach to this problem involved `Third Quantization' and `baby universes', following an idea of Sidney Coleman \cite{Coleman:1988tj}.} Explaining why the cosmological constant $\Lambda$ vanishes was considered to be the greatest puzzle for theoretical physics and all previous approaches to explain it had failed.

In preparing for the lectures, Weinberg addressed the question in his characteristically systematic way, with several original variations \cite{Weinberg:1988cp}. While describing attempts to use approximate scale invariance to solve the problem he came up with a no-go theorem, now known as {\it Weinberg's theorem} explaining why scale invariance in itself cannot solve the problem. He also refined his earlier analysis of the merits of an anthropic explanation to the cosmological constant \cite{Weinberg:1987dv}, that argue that the universal expansion must be consistent with the formation of large-scale structures (like galaxies) since so far as we know these seem to be required for the existence of life, and so also for the existence of the humans for whom the problem seems to be so important. His key observation was that if this line of argument is valid then there should also be a prediction: the cosmological constant is likely to be relatively close to its allowed upper limit.  

This was a very brave article written more than 10 years before experiments found evidence for what is now called Dark Energy, whose density is indeed in the ballpark Weinberg predicted. His argument, as extended by Raphael Bousso and Joseph Polchinski \cite{bousso} in the context of the `landscape' of possible extra-dimensional geometries allowed by string theory, is one of the most concrete approaches to the Dark Energy so far on the table (though the last word remains to be said).

\item{\bf Experimental tests of quantum mechanics}

In the late 1980s Weinberg returned to the foundations of quantum mechanics (QM), in an effort to find a class of generalizations of QM that could be compared with observations, if only to be better able to quantify the accuracy with which QM prevails. He was painfully aware when doing so that foundational issues in quantum mechanics are famously the hill on which elder physicists go to die, so insisted he was not yet old enough for that and kept his focus on potential experimental tests of quantum mechanics itself (most notably tests of non-linear modifications of the linear principle of superposition). His work stimulated a revival of precision experiments to test these ideas, with (perhaps needless to say) quantum mechanics emerging triumphant.
\end{itemize}

\subsection{1990s-2020s}

During his last 30 years, after entering his 60s, Weinberg's intellectual energy began to be funnelled more into other activities besides research, including finishing 7 of his 8 textbooks and 6 of his 8 popular science books. Nevertheless he still kept a Weinbergian weather eye on research, publishing research papers right up until shortly before his death. We highlight the following of his contributions during these twilight decades:

\begin{itemize}
\item {\bf Nuclear forces from chiral lagrangians}

In the 1990s Weinberg returned to his roots, applying his earlier insights into low-energy pion physics to study nuclear physics in terms of effective field theories. This was his way of creatively taking on board developments at the time incorporating nonrelativistic physics into the EFT framework, something that is possible because the expansion in powers of a slow speed $v \ll 1$ can also be interpreted as an expansion in a hierarchy of scales: $E \simeq \frac12 mv^2 \ll p \sim mv \ll m$. The utility of applying EFT methods to nonrelativistic expansions in this way was becoming appreciated around this time in the context of understanding the decays of mesons involving heavy quarks (like the $c$ or $b$ quark) \cite{Isgur:1989vq}, generalizing earlier work that had been done for precision calculations of radiative corrections to atomic bound states \cite{Caswell:1985ui} 

Weinberg used similar tools to compute the inter-nucleon forces implied at low energies by generalizing the effective theory governing low-energy pion interactions to include nonrelativistic nucleons \cite{Weinberg:1990rz, Weinberg:1991um,Weinberg:1992yk} (see also \cite{Ordonez:1995rz}). By so doing he enabled the calculation of \emph{ab initio} nuclear energy levels for the first time, at least for light nuclei involving comparatively few protons and neutrons. Nuclear physicists at the time were instead fitting data from nuclear measurements with models in which meson exchange between nucleons assumed phenomenological couplings. The new conceptual feature was the systematic use of the most general couplings allowed by the symmetries of QCD at low energies, since this ensures that the result must capture the low-energy limit of QCD itself (as opposed to `just' being a more heuristic model). It is fair to say that his ground-breaking papers \cite{Weinberg:1990rz, Weinberg:1991um} -- which have to date accrued more than 1500 citations each -- opened a new approach towards theoretical nuclear physics\footnote{For a recent essay describing the history and relevance of this initiative see \cite{vanKolck:2021rqu}.}.

\item{\bf Cosmology rebooted}

By the mid-1990s the Standard Model had been tested at the per-mille level, such as through measurements of high-energy $e^+e^-$ collisions at the LEP facility at CERN in Geneva, and had passed the test spectacularly. By contrast, precision measurements of the Cosmic Microwave Background were about to bear fruit and possibly bring some order to the Wild West of cosmological model building. Given his attachment to theoretical questions immediately related to experiments, Weinberg (like many particle physicists at the time) turned his attention towards cosmology. In particular he devoted considerable effort to understanding from first principles the theory of cosmological perturbations, that governs the observables being studied in the CMB.  

The high hopes for cosmology were confirmed by several rapid discoveries in the 1990s: the measurement of the fluctuations in the CMB temperature and the discovery that the universal expansion is accelerating rather than decelerating stimulated the emergence of the $\Lambda$CDM `cosmological standard model', which provides a good fit to a great many measurements using only a handful of parameters. Intriguingly, it does so only if the present-day universal energy density contains two unknown sources of energy; the so-called Dark Matter and Dark Energy.

Besides expanding and solidifying his work on the cosmological constant from the 1980s, Weinberg threw himself into recalculating the evolution of perturbations in General Relativity relevant to understanding the temperature fluctuations in the CMB. This work often confirmed results made earlier by cosmologists themselves, for whom the evolution of cosmic perturbations had a much longer history \cite{Mukhanov:1981xt, Hawking:1982cz, Guth:1982ec,Linde:1982uu, Bardeen:1983qw, Brandenberger:1983tg, Lyth:1984gv} but usually from a different perspective and with an eye on the generality of the conclusions. Two highlights from this time are:

\begin{enumerate}
\item {\bf Adiabatic modes in cosmology:} Weinberg proved rigorously the conservation of super-Hubble comoving curvature perturbations during and after inflation \cite{Weinberg:2003sw}, confirming a result previously known but under very general circumstances. The importance of this conservation is that it allows the spectrum of perturbations to be evolved from inflation to the much later observable universe in a largely model-independent way that does not depend on the may complicated details of the intervening evolution, which could well be very complicated and involve poorly understood physics. This is a key result that makes the predictions of inflationary models for the later universe much more robust than one might have expected.

\item  {\bf Quantum corrections to cosmological correlators:} Weinberg set up the systematic framework for the study of cosmological correlators to all orders in a perturbative expansion \cite{Weinberg:2005vy}, building on earlier calculations of non-gaussianities \cite{Maldacena:2002vr}. In particular he popularized the more widespread use of the in-in formalism -- originally introduced by Schwinger and Keldysh \cite{Schwinger:1960ybt, Keldysh:1964ud} (as opposed to the more familiar in-out formalism that arises in scattering theory) -- in cosmological circles. This established what is now the standard approach used by theoretical cosmologists. Along the way he showed that loop corrections to cosmological correlators can be done reliably, arguing this to be worthwhile even if they turn out too small to be detectable. In his own words: 

\begin{quote}
{\it ...just as field theorists in the 1940s and 1950s took pains to understand quantum electrodynamics to all orders in perturbation theory, even though it was only possible to verify results in the first few orders. }
\end{quote}


\end{enumerate}

\end{itemize}

To these could be added many other subjects to which Weinberg brought his special touch. Sadly, the great breadth of his contributions, combined with the very high bar set by his most important results, makes their complete discussion beyond the scope of this article.


\section{Wider Writings}

As his impact in science grew so did his prestige, both amongst scientists and lay people. This section is devoted to a short summary of his non-research writing and other activities.

\subsection{Essays}

Over the years Weinberg was often invited to speak in more general settings and to contribute to volumes commemorating other scientists.\footnote{Weinberg was never the subject of similarly celebratory conferences during his lifetime because he believed them to be poor uses of research funds. } His talks were often gems because he would use them to provide the Big Picture, as he saw it, and how it had evolved during his lifetime. His words on such occasions were often recorded and later released as essays for a wider readership. Although they often do not describe original research they are full of Weinbergian wisdom: clarity of thought conveyed concisely with a touch of wry humour. This section briefly records our favourite selection of these.

\begin{itemize}

\item{\bf Why the renormalization group is a good thing.}

This article \cite{RGGoodThing} appears in a book written to commemorate the famous Gell-Mann and Low article defining the renormalization group (RG). He emphasises this article's importance for understanding the running of gauge couplings and provides one of the most lucid accounts of how RG methods can be used to extend the reach of perturbative methods by resuming the leading logarithms -- {\it i.e.}~working to all orders in $\alpha \log(E_1/E_2)$ but neglecting higher powers like $\alpha^2 \log(E_1/E_2)$, where $\alpha \ll 1$ is a small perturbative coupling while $E_1$ and $E_2$ are two very different energy scales. 

Along the way he also playfully introduces and explains his Three Laws of Theoretical Physics:
\begin{quote}{\bf Weinberg's Laws of Progress in Theoretical Physics}
\item {\sl 1. You will get nowhere by churning equations (conservation of information).}
\item  {\sl 2. Do not trust arguments based on the lowest  order of perturbation theory.}
\item {\sl 3. You may use any degrees of freedom you like to describe a physical system, but if you use the wrong ones, you'll be sorry. }
\end{quote}

\item{\bf Superconductivity for particular theorists}

In this essay \cite{Weinberg:1986cq} -- written on the occasion of one of Yoichiro Nambu's birthdays -- Weinberg wrote a beautiful illustration of the power of effective field theories used in an unfamiliar setting (for particle physicists). His vehicle for doing so is to explain to others something that condensed matter physicists had discovered long ago: how some properties of superconductors -- such as the frequency of the Josephson effect -- can be predicted with much more accuracy than seems like should be allowed by the accuracy of the underlying models of the superconducting mechanism itself (such as the Bardeen-Cooper-Schrieffer model of pairing).  

The point is that the underlying models aim to explain why the ground state of a superconductor spontaneously breaks the electromagnetic $U_\EM(1)$ gauge group, but once this breaking is achieved some of the predictions can be extremely accurate because they are simply consequences of the symmetry breaking pattern and do not depend at all on the precise mechanism that accomplishes it. In the superconducting case these model-independent predictions are the ones that describe the interactions of the would-be Goldstone mode for this symmetry breaking. Using the EFT of this mode Weinberg shows how all of the main properties of superconductors follow: vanishing resistance, the Meissner effect, the Josephson frequency and so on. The result is vintage Weinberg, explaining many physical properties from a simple principle generally applied. 

\item{\bf What is quantum field theory, and what did we think it is?}

In this lovely essay \cite{Weinberg:1996kw} Weinberg describes the long arc of the development of quantum field theory and how this has influenced our picture of what the fundamental constituents of nature are. The story starts with an initial picture of classical particles interacting with classical fields, all of which get `quantized' once quantum mechanics is discovered. In the standard narration the picture then evolves towards a framework in which everything is a field, with particles being just their quanta of excitation. 

Weinberg's telling of this story broadly agrees with this general arc, but is also interestingly different. He describes the `folk theorem' that there is no content in relativistic quantum field theory beyond the principles of Lorentz invariance, quantum mechanics and a third principle: `cluster decomposition'. Cluster decomposition is the statement that quantum amplitudes that are widely separated in space at a given time should remain factorized if they initially were factorized. Factorization is what would normally be expected for statistically independent events and plays an important role in the phenomenon of spontaneous symmetry breaking. It is also ultimately why locality plays an important role. 

In Weinberg's picture one does not take for granted why one studies quantum field theory. Instead he argues that by using quantum field theory we are only building in these fundamental principles (plus the presence of a finite number of degrees of freedom, such is usually true at low energies even if not true at all energies, {\it e.g.}~in string theory). This was also the way he taught QFT: in what was typically a two-year cycle topics that are often the starting point in other approaches to the subject (like canonical methods, lagrangian formulations or path integrals) were not introduced until the need for them is first justified, often not until after January in the first year.

\item{\bf Advice to young scientists}

In June 2003 Weinberg was awarded an honorary doctoral degree by the Science Faculty at McGill University. In his commencement address he chose to give career advice to young scientists who were just starting out, something he had not done when receiving similar honours elsewhere because in those earlier occasions the convocations were shared with engineering graduates. 

His advice in a nutshell: 
\begin{enumerate}
\item  {\it No one knows everything and you don't have to.} 
\item   {\it Seek out where things look messy -- that's where the action likely is.}
\item  {\it Forgive yourself for wasting time.} 
\item {\it Learn something about the history of science...you can get great satisfaction by recognizing that your work in science is a part of a grand sweep of history. But also because the best antidote to the philosophy of science is a knowledge of the history of science.}
\end{enumerate}
His advice was later recorded as an essay `Four Golden Lessons' \cite{GoldenLessons}, and contains many insightful and irreverent observations about how to choose research topics to best make an individual impact on science that remain a must-read for scientists both young and old. 

His address closes with the importance of remembering the impact science has made on mankind's overall intellectual bearings, using as an example how the scientific determination of the age of the Earth removed the intellectual respectability of a belief in the Bible as a literal description of the world. In a master-stroke of timing, his address was immediately followed in the convocation ceremony by the benediction, in which a clergy member blessed the proceedings.

\end{itemize}

\subsection{Textbooks}

Weinberg's textbooks are simply classics in their respective fields. They reflect his unique ability to present things straightfowardly, but also as generally as possible. They vary in depth and mathematical sophistication. He was not afraid of using mathematical tools if necessary, but he did not necessarily use mathematical sophistication if he thought it superfluous. 

\subsubsection{Gravitation and Cosmology}

Weinberg's first textbook  \cite{Weinberg:1972kfs} was in a field that was not his main research subject. In 1972 he wrote the extremely influential book {\it Gravitation and Cosmology: Principles and Applications of the General Theory of Relativity}. His presentation is clear and direct, possibly because his learning of the subject was fresh in his mind. Its great strength is its focus on physical applications, unlike the trend amongst professional relativists who tended to insist on more abstract aspects of the mathematics. This book is about things falling, rather than about fibre bundles. He gloried in being explicit, including in his treatment of tensorial indices. His notation is widely used, and although hated by some it is loved by all right-thinking people.

The book covers the basic mathematics needed to understand the geometry of General Relativity but takes a particle physicist perspective that also tries to emphasize the similarities with other forces found in nature. It starts with general aspects of Special Relativity, extends to the case of General Relativity and includes concrete solutions of Einstein s equations such as the Schwarzschild and Friedmann-Lemaitre-Robertson-Walker cosmology. In an interesting omission he mentions only occasionally the word black hole even though he briefly talks about gravitational collapse.  On the other hand he dedicates great effort to gravitational waves and the thermal history of the universe as well as post-Newtonian approximations.

His presentation of cosmology was highly influential because at that time most high-energy physicists were unfamiliar with cosmology and the concepts and calculational tools of astrophysics. It still reads in a very modern way, despite being written more than 50 years ago. Its treatment is out of date on some topics, such as the growth of fluctuations within expanding universes, something he corrected in his later books. But as a general introduction to the subject it remains one of the best entry-level treatments of General Relativity and its applications.

\subsubsection{Quantum Theory of Fields: Volumes 1,2,3}

Quantum field theory is arguably Weinberg's core competency and he was involved in many of the developments that made it into a central pillar of modern physics. His three-volume magnum opus \cite{Weinberg:1995mt, Weinberg:1996kr, Weinberg:2000cr} on the subject provides arguably the deepest insights and the most definitive treatment of the field, covering many topics not broached in other textbooks. The books were based on his lectures and many decades of thinking about the subject's foundations, and partly because of this it is a harder read for those just learning the field than is Gravitation and Cosmology. Like great religious texts one reads Weinberg's QFT volumes multiple times over one's life and learns something new every single time.

The organisation, especially of the first volume is non-standard (and not completely user-friendly) but follows his own way of understanding the field. The following  points are emphasised throughout: 
\begin{itemize}
\item From particles to fields. He starts with a complete description of the symmetries of Special Relativity and -- following the work of Wigner in the 1930s and 1940s -- introduces particles as unitary representations of the Poincare group, labelled by a few quantum numbers such as mass, spin, momenta and third component of angular momentum or spin. He shows how consistency between Quantum Mechanics and Special Relativity is delicate: the consistency of things like time-ordering (as appears in perturbative expressions for scattering amplitudes) with relativity suggesting that interaction Hamiltonians should be local and so be expressed in terms of quantum field operators regarded as functions of spacetime position.   In his own words:
\begin{quote}
{\it  If it turned out that some physical system could not be described by a quantum field theory, it would be a sensation; if it turned out that the system did not obey the rules of quantum mechanics and relativity, it would be  a cataclysm.}
\end{quote}

\item  Effective field theories. His books reflect his deep understanding of effective field theories. In the sense that the organisation of the books is done with the vision of EFTs playing a central role. In his own words:
\begin{quote}
{\it This book is written for the effective field theory era...}
\end{quote}
\item The discussion of lagrangians to describe interactions is introduced only in chapter 7, after the implications of representations of the Poincare group are studied in detail. The motivation for introducing the lagrangian and using canonical quantization is that it provides a systematic way to generate the non-Lorentz-invariant interactions into the interaction Hamiltonian that are often required to cancel Schwinger terms in time-ordered correlation functions. Path integrals are later motivated as a way to avoid using Hamiltonians altogether, allowing things to be kept manifestly covariant at every step.

\item Foundational issues are given full attention. For instance, the description of how the gauge redundancy arises in the field description of massless helicity $\pm 1$ states as a consequence of Lorentz invariance and general quantum principles is made explicitly. He also has a general discussion of soft theorems as part of his discussion of infrared behaviour in QED, following his own research. General results such as the detailed study of $C,P,T$ symmetries, the $CPT$ theorem, the spin-statistics connection, the Goldstone theorem, the Coleman-Mandula theorem, the limit on the number of supersymmetries, \emph{etc.}~are discussed systematically. All of these ingredients make these books stand out from other QFT textbooks. 
\end{itemize}




\subsubsection{Cosmology}

In 1999 -- after finishing his QFT books -- Weinberg decided to write a book about cosmology \cite{Weinberg:2008zzc}. The field had developed so much since his earlier book on gravitation and cosmology that a totally new book was needed. The book covers modern cosmology especially emphasising the (model independent) physics of inflation and the perturbations of the cosmic microwave background.  In preparing the book he redid from scratch most of the calculations that professional cosmologists had been doing over the years, often based on computer simulations. With his typical strong stomach for calculating he managed to reduce most calculations to the point where the main physical issues were transparent without relying on numerics.

\subsubsection{Lectures on Quantum Mechanics}

This was a highly anticipated book \cite{Weinberg:2013}. Weinberg adopted the tradition of writing a book on the courses he was teaching at the University of Texas. This book emerged as the result of several semesters of lecturing Quantum Mechanics for advanced undergraduate students. It covers the basics of Quantum Mechanics but unlike older books it gives more emphasis to timely subjects like entanglement entropy, Bell inequalities,  Berry's phase, the in-in formalism, Dirac's constrained canonical systems, open systems, etc. 

A topic deliberately \emph{not} covered is the frequently made argument that the Dirac equation is the relativistic generalisation of Schr\"odinger's equation. Weinberg regarded this historical approach to be at best misleading. There should not be a unique relativistic generalization of the Schr\"odinger equation since the Dirac equation applies only to spin-half particles, and other spins are possible with the corresponding fields having different equations. 

Finally, the book summarizes the various proposals for the interpretation issues of Quantum Mechanics; a most welcome complete and up-to-date snapshot of this field. His telling of the story finds all of the proposed interpretations to be wanting, but he takes his time to show why he believed this is true.

\subsubsection{Lectures on Astrophysics}

This small monograph \cite{Weinberg:2019} was also based on a course he taught at UT and -- much like for his Cosmology book -- Weinberg included much original material since he insisted in reproducing calculations in his own way, mostly using analytical rather than numerical methods. This makes the physical interpretation of his results often more transparent than more standard treatments. The book covers the foundations of astrophysics (such as the physics of the large scale structure of the universe including stars, galaxies and interstellar medium) and the interplay between astrophysics and high energy physics is emphasised, including a detailed description of possible sources of gravitational waves.

\subsubsection{Foundations of Modern Physics}

His last textbook \cite{Weinberg:2021kzu} aims to introduce undergraduates to modern physics. It includes many topics -- like thermodynamics and statistical mechanics -- that were not as familiar to him as other subjects,\footnote{Weinberg confessed in private conversations that he particularly enjoyed  teaching thermodynamics, a subject he had never taught before, doing so for  first time at age 83.} 
but also includes more straightforward presentations of Special Relativity, Quantum Mechanics and even the elements of QFT. It is a very good book to give students a general overview of modern physics before starting graduate school.

\subsection{Popular Science Books}

One quality that separates Weinberg from most leading theoretical physicists was his role as public intellectual. Full of wisdom in the broader sense his wide reading gave him an encyclopedic knowledge of history and other human affairs. His talent for weaving erudition with common sense makes his public writings both pleasing and informative to read. His eight influential popular science books have inspired several generations and leave a lasting legacy.

It is impossible to summarise all of his writings in detail. We content ourselves here with brief overviews of the content of the books, supplemented by selected quotes that best convey his gift for capturing the gist of a thing in a few artfully chosen words. You can sometimes almost hear the characteristic New York accent.

\subsubsection{The First Three Minutes: A Modern View of the Origin of the Universe}

This highly influential book  \cite{Weinberg:1977ji} was written in the 1970s and has been a standard reference for lay people interested in understanding the science behind Big Bang cosmology. Written for the general public with only a few equations, it describes the physical principles and the history of the major discoveries that led to the standard model of cosmology. From the abundance of the different elements to the cosmic microwave background that confirmed the validity of the Big Bang scenario and ruled out alternative proposals such as the steady state theory.

Even though the book was written for the general public it was often also a natural source for professional high energy theorists to learn the basics of cosmology at a time when cosmology was not a mainstream subject of study for particle physicists. The revised edition in the 1990s updated the original book to include ideas of the inflationary universe, started by Alan Guth (who acknowledges having learnt cosmology from the first edition of this book) and others. Even though other books are more up to date, this book remains as a classic in the subject. 

Selected quotes from this book:
\begin{itemize}

\item 
On the importance of the discovery of the cosmic microwave background (CMB):
\begin{quote}
{\it The most important thing accomplished by the ultimate discovery of the 3° K radiation background (Penzias and Wilson, 1965) was to force all of us to take seriously the idea that there was an early universe.}
\end{quote}

\item On the importance of trying to understand the universe
\begin{quote}
{\it The effort to understand the universe is one of the very few things that lifts human life a little above the level of farce, and gives it some of the grace of tragedy.}
\end{quote}

\item An honest and difficult conclusion of our current understanding of the universe 
\begin{quote}
{\it The more the universe seems comprehensible, the more it also seems pointless.}
\end{quote}

\item An optimistic view of the satisfaction of the research endeavour
\begin{quote}
{\it If there is no solace in the fruits of our research, there is at least some consolation in the research itself. Men and women are not content to comfort themselves with tales of gods and giants, or to confine their thoughts to the daily affairs of life; they also build telescopes and satellites and accelerators and sit at their desks for endless hours working out the meaning of the data they gather.\footnote{Years later on a PBS interview \cite{pbs} Weinberg expanded on this as follows: {\it ...  if there is no point in the universe that we discover by the methods of science, there is a point that we can give the universe by the way we live, by loving each other, by discovering things about nature, by creating works of art. And that -- in a way, although we are not the stars in a cosmic drama, if the only drama we're starring in is one that we are making up as we go along, it is not entirely ignoble that faced with this unloving, impersonal universe we make a little island of warmth and love and science and art for ourselves. That's not an entirely despicable role for us to play.}}}
\end{quote}

\item  A lesson to theorists
\begin{quote}
{\it This is often the way it is in physics. Our mistake is not that we take our theories too seriously, but that we do not take them seriously enough. It is always hard to realize that these numbers and equations we play with at our desks have something to do with the real world.}
\end{quote}
\end{itemize}

\subsubsection{The Discovery of Subatomic Particles}

In this special book  \cite{Weinberg:1983}, written for a Scientific American collection, Weinberg illustrates his passion for the history of science. He describes the main theoretical and experimental discoveries that have led towards the current understanding of the fundamental building blocks of matter. From the discovery of the electron, the structure of the atom with the nucleus of Hydrogen, the proton, as a second building block and the neutron as the third to the myriad of elementary particles discovered in the 1950s and 1960s to the current order in terms of a few quarks and leptons. 

The book nicely illustrates the different aspects of scientific research and scientific discoveries that usually take many turns before converging to a proper understanding of this field. It is a very valuable source on the 20th century history of science.  

\begin{itemize}
\item On his own knowledge of history of particle physics:
\begin{quote}
{\it My hope… is that this book may contribute to a radical revision in the way that science is brought to nonscientists… This book is intended to be comprehensible to readers who have no prior background in science, and no familiarity with mathematics beyond arithmetic… Although this book is written for the nonscientist, it has one aspect that perhaps also my fellow physicists may find interesting. The great scientific achievements described here form a large part of the soil from which our own more recent harvest of discoveries have sprung. Yet I, for one, had only the foggiest idea of the early history of twentieth-century physics when I started to teach the courses at Harvard and Texas, and I suspect that the same is true of many of my colleagues in physics}
\end{quote}
\item
On theorists and experimentalists
\begin{quote}
{\it ...during the exciting period in the 1920s when quantum mechanics was being developed, a colleague asked, ‘How is physics these days, [Ernest] Rutherford?’ and Rutherford replied, ‘…the theorists are on their hind legs and it is up to us to get them down again.’ As a theorist I naturally tend to deplore this sort of anti-theoretical feeling. But in fact theorists and experimentalists generally get along pretty well with each other, and could hardly get along at all without each other.}
\end{quote}

\item
On the importance of particles
\begin{quote}
{\it The real task we address … is not to develop a list of particles and their properties. It is to understand the underlying principles that dictate why nature… is the way it is. All our experience shows that the study of elementary particles is at present the best and perhaps the only way of getting at the fundamental laws of nature.}
\end{quote}
\end{itemize}

\subsubsection{Dirac Lectures: Elementary Particles and the laws of Physics}

To honour the late Paul Dirac, the Department of Applied Mathematics and Theoretical Physics (DAMTP) of the University of Cambridge decided to establish a prestigious yearly set of lectures named after him. Due to some organization complication the first two Dirac lectures were given the same year (1985), by Richard Feynman and Steven Weinberg. DAMTP decided to combine them into one single  book  \cite{Feynman:1987gs}. 

Being lectures for a general physics audience, the level is higher than a standard popular science book (including some equations). But it is a good sample of the thinking of two of the brightest minds of the 20th century.

Weinberg's part was mostly about the dreams of a final theory (more about which below).

\subsubsection{Dreams of a Final Theory: The Search for the Fundamental Laws of Nature}

This book  \cite{Weinberg:1992nd} is a rare masterpiece that most probably will be quoted for years to come. It is a definitive reference for future scientists and philosophers of science about what scientists in the end of the 20th century thought about the importance of science itself (and its important open questions). 

The book grew out of Weinberg's ultimately unsuccessful campaign in support of building the Superconducting Super Collider (SSC); an ambitious project of the 1990s to build the ultimate particle collider. In the end his efforts and those of other members of the high energy physics community were in vain because the US congress chose to stop the project. Europe eventually built a somewhat smaller version -- the LHC (Large Hadron Collider) -- about a decade later, in which the Higgs particle was eventually discovered. 
The SSC being more ambitious, probing twice the energy of the LHC, may have discovered the Higgs earlier and possibly other particles not yet seen at the LHC. A small consolation of Weinberg's lost effort was the offering to the world of this exceptional book.

\begin{itemize}
\item On teaching physics
\begin{quote}
{\it I have felt that my most important task (and certainly the most difficult) was to give the students a taste of the power of being able to calculate in detail what happens under various circumstances in various physical systems. They were taught to calculate the deflection of a cathode ray or the fall of an oil droplet, not because that is the sort of thing everyone needs to calculate but because in doing these calculations they could experience for themselves what the principles of physics really mean. }
\end{quote}

\item On reductionism
\begin{quote}
{\it Once again I repeat: the aim of physics at its most fundamental level is not just to describe the world but to explain why it is the way it is.}

\end{quote}

\item On beautiful theories
\begin{quote}
{\it Plato and the neo-Platonists taught that the beauty we see in nature is a reflection of the beauty of the ultimate, the nous. For us, too, the beauty of present theories is an anticipation, a premonition, of the beauty of the final theory. And in any case, we would not accept any theory as final unless it were beautiful.}
\end{quote}

\pagebreak
\item On the Final Theory
\begin{quote}
{\it The dream of a final theory inspires much of today’s work in high-energy physics, and though we do not know what the final laws might be or how many years will pass before they are discovered, already in today’s theories we think we are beginning to catch glimpses of the outlines of a final theory.}
\end{quote}

\item On God
\begin{quote}
{\it The theologian Paul Tillich once observed that among scientists only physicists seem capable of using the word `God' without embarrassment. Whatever one's religion or lack of it, it is an irresistible metaphor to speak of the final laws of nature in terms of the mind of God.}
\end{quote}

\item On Philosophy
\begin{quote}
{\it The value today of philosophy to physics seems to me to be something like the value of early nation-states to their peoples. It is only a small exaggeration to say that, until the introduction of the post office, the chief service of nation-states was to protect their peoples from other nation-states. The insights of philosophers have occasionally benefited physicists, but generally in a negative fashion—by protecting them from the preconceptions of other philosophers.}
\end{quote}
\end{itemize}

\subsubsection{Facing-Up: Science and its Cultural Adversaries}

This book \cite{facing} is a collection of essays about the place of science within different intellectual activities. In his own words he  enjoyed a sense of controversy and wrote the different essays based on talks and interviews he was invited to give over several years. The title is inspired by a statue of the astronomer Tycho Brahe in a posture of facing up that he interpreted as being about the scientists facing up to their understanding of a world without a concrete purpose and no special role for humans.

As in his previous books he promotes the scientific point of view regarding reductionism and rejects other views based on philosophy and religion. 
In particular he became involved in the `science wars' initiated by the famous article submitted to a social science journal by NYU Professor Alan Sokal. The article was a made-up concoction of non-sensical ideas claimed to be related to physics that was meant to highlight a lack of rigour in the social sciences. It was accepted for publication. This so-called `Sokal's hoax'  created a back and forth discussion among social and quantitative scientists and naturally attracted Weinberg's attention. He wrote several articles about it.

Some selected quotes:
\begin{itemize}
\item On the importance of great research universities:
\begin{quote}
{\it I am convinced that without great research universities we in the United States would have to support ourselves by growing soy beans and showing the Grand Canyon to tourists from Germany and Japan.}
\end{quote}

\pagebreak
\item On the importance of reductionism:
\begin{quote}
{\it There are arrows of scientific explanation, which thread through the space of all scientific generalizations. Having discovered many of these arrows, we can look at a pattern that has emerged, and we notice a remarkable thing: perhaps the greatest scientific discovery of all. These arrows seem to converge to a common source! Start anywhere in science and, like an unpleasant child keep asking ``Why?'' You will eventually get down to the level of the very small.}
\end{quote}

\item On his own quote from the first three minutes:
\begin{quote}
{\it But the tragedy is not in the script; the tragedy is that there is no script.}
\end{quote}

\item On string theory:
\begin{quote}
{\it It is because we expect that string theory will be testable -- if not directly by observing the string vibrations, then indirectly, by calculating whether string theory correctly accounts for all of the currently mysterious features of the Standard Model of elementary particles and General Relativity. If it were not for this expectation, string theory would not be worth  bothering with. }
\end{quote}
\item On the Science Wars: 
\begin{quote}
{\it Endless trouble has been produced throughout history by the effort to draw moral or cultural lessons from discoveries of science.}
\end{quote}
\item On some trends in the social sciences:
\begin{quote}
{\it In trying to get votes for the Superconducting Super Collider, I was very much involved in lobbying members of Congress, testifying to them, bothering them, and I never heard any of them talk about postmodernism or social constructivism. You have to be very learned to be that wrong. }
\end{quote}
\item Criticizing the philosophy of de-constructivist philosopher Derrida:
\begin{quote}
{\it It seems to me that Derrida in context is even worse than Derrida out of context.}
\end{quote}
\item On his disagreement with ``The Structure of Scientific Revolutions'', by Thomas Kuhn:
\begin{quote}
{\it What does bother me on rereading Structure and some of Kuhn's later writings is his radically skeptical conclusions about what is accomplished in the work of science. And it is just these conclusions that have made Kuhn a hero to the philosophers, historians, sociologists, and cultural critics who question the objective character of scientific knowledge, and who prefer to describe scientific theories as social constructions, not so different in this respect from democracy or baseball. }
\end{quote}
\end{itemize}

\subsubsection{Lake Views: This World and the Universe}

This is also a collection of essays  \cite{views} written on different occasions; often published in the New York Review of Books and other repositories for popular science articles. They cover his thoughts on general scientific issues, following up his discussions on his previous books. In them he emphasizes his scientific rational perspective, including cosmology and reductionism, as the basis of understanding science, but also discusses several more worldly issues such as his strong rejection of nuclear weapons, his support for the state of Israel, arguments against sending humans to space, advice to young scientists (the advice listed above), and more.

Selected quotes from this book:

\begin{itemize}

\item
On then-proposed missile-defense systems
\begin{quote}
{\it If it were possible tomorrow to switch on a missile defense system that would make the United States invulnerable to any missile attack, then I and most other opponents of missile defense would be all for it. But that is not  the choice we face.}
\end{quote}
\item
On the danger of nuclear weapons
\begin{quote}
{\it It should be clear by now that national security is not always best served by building the best weapons.}
\end{quote}
\item On the Universe as a Computer (reviewing Stephen Wolfram's ``A New kind of Science''):
\begin{quote}
{\it ... he concludes that the universe itself would then be an automaton, like a giant computer. It's possible, but I can't see any motivation for these speculations, except that this is the sort of system that Wolfram and others have become used to in their work on computers. So might a carpenter, looking at the moon, suppose that it is made of wood.}
\end{quote}
\item 
On the cultural importance of radioactivity:
\begin{quote}
{\it How important is now who was prime minister of Great Britain or Canada in 1903? What stands out as really important is that at McGill University, Ernest Rutherford and Frederick Soddy were working out the nature of radioactivity...The understanding of radioactivity allowed physicists to explain how the Sun and Earth's cores could still be hot after millions of years. In this way, it removed the last scientific objection to what many geologists and paleontologists thought was the great age of the Earth and the Sun. After this, Christians and Jews either had to give up belief in the literal truth of the Bible or resign themselves to intellectual irrelevance.}
\end{quote}
\item
On Einstein's mistakes:
\begin{quote}
{\it Occam's razor is a fine tool, but it should be applied to principles, not equations.}
\end{quote}
\begin{quote}
{\it It is enough to say that neither Bohr nor Einstein had focused on the real problem with quantum mechanics...The difficulty is not that quantum mechanics is probabilistic -- that is something we apparently just have to live with. The real difficulty is that it is also deterministic, or more precisely, that it combines a probabilistic interpretation with deterministic dynamics.}
\end{quote}
\begin{quote}
{\it Perhaps Einstein's greatest mistake was that he became the prisoner of his own successes. }
\end{quote}

\item 
On his belief in the multiverse:
\begin{quote}
{\it  Martin Rees said that he was sufficiently confident about the multiverse to bet his dog's life on it, while Andrei Linde said he would bet his own life. As for me, I have just enough confidence about the multiverse to bet the lives of both Andrei Linde and Martin Rees's dog.}
\end{quote}
\end{itemize}

\subsubsection{To Explain the World: The Discovery of Modern Science}

This book  \cite{explain} shows Weinberg at his best. It describes events in the history of modern science as seen by a scientist rather than a historian. The standard attitude in history is not to judge historical events based on our current vision, instead attempting to see them through the vision of that time. Imposing a narrative in which historical events are part of an inevitable journey towards present-day values are called `whig history', a term historian Herbert Butterfield introduced in 1931. In this book Weinberg makes the case for what he calls the whig history of science. He argues we can actually judge events in the history {\it of science} using the objective criterion of what turned out to be correct and what did not.  

As might have been expected, the book created controversy (to Weinberg's delight) and received mixed reviews. The content is fascinating: an assessment of the importance of developments in the history of the physical sciences by probably the world's most qualified person to do so. It covers most western developments emphasising contributions from the Greeks up until the 19th century, including some contributions from Asia and the Middle East, though in less detail  (see also \cite{physicsworld}). He includes some reflections about our current times at the end. With Weinbergian relish, he is not afraid to dive into calculational details in old theories, such as explaining the calculations that led Ptolemy to his theory about the motion of celestial objects with  Earth as the center. He remarks that not being intimate with the sea did not preclude Newton from being able to explain the tides from the gravitational effect of the moon. A wonderful read.
\\

Some quotes from this book:

\begin{itemize}

\item 
On pre-scientific disciplines
 \begin{quote}
{\it Indeed, none of the early Greeks from Thales to Plato, in either Miletus or Abdera or Elea or Athens, ever took it on themselves to explain in detail how their theories about ultimate reality accounted for the appearances of things.}
\end{quote}

\item Criticism of Aristotle and Plato in one sentence
\begin{quote}
{\it ...but although often wrong Aristotle is not silly, in the way that Plato sometimes is.}
\end{quote}
\item Separating science and religion
\begin{quote}
{\it It was essential for the discovery of science that religious ideas be divorced from the study of nature. }
\end{quote}
\item Importance of asking relevant questions
\begin{quote}
{\it The progress of science has been largely a matter of discovering what questions should be asked.}
\end{quote}
\item On the negative reactions against the anthropic principle:
\begin{quote}
{\it Whatever the final laws of nature may be, there is no reason to suppose that they are designed to make physicists happy.}
\end{quote}
\end{itemize}

\subsubsection{Third Thoughts}

Weinberg's third collection of essays  \cite{third}, following on Facing Up and Lake Views, expands on similar themes as in his previews books. It comes in four parts: Science History, Physics and Cosmology and Public and Private  Matters. Again, he is not afraid of making strong claims on his views regarding science and its history but also on matters such as deciding not to vote for any presidential candidate in the 2012 election, his controversial comparison between the crafts of science and art or against sending manned missions to space.

Some quotes from this book:\footnote{See also Luciano Maiani's lovely essay about Weinberg \cite{luciano}.}

\begin{itemize}

\item
A private farewell?
\begin{quote}
{\it From past experience, it seems that at my rate of writing it takes about a decade to produce enough new essays for assembly in a collection. I hope nevertheless that this will not be my last collection. But given actuarial realities, perhaps this would be a good time for me to add a word of thanks to readers who over many years have put up with my polemics and explanations, and have thereby given me a precious contact with the world beyond physics.}
 \end{quote}
\item
On the Crisis of Big Science
\begin{quote}
{\it My pessimism comes partly from my experience in the 1980s and 1990s in trying to get funding for another large accelerator...I gather that dollar for dollar, government spending stimulates the economy more than tax cuts. It is simply a fallacy to say that we cannot afford increased government spending. But given the anti-tax mania that seems to be gripping the public, views like these are political poison. This is the real crisis, and not just for science.}
 \end{quote}
\item
On liberal disappointment
\begin{quote}
{\it I will just express my opinion that if Obama had vigorously taken up the issue of inequality, and in particular had espoused the cause of organized labor, then, even if he had not had his way with Congress, he would have raised a banner around which liberals and working people could gather, and we might not  now have to endure the presidency of Donald Trump.}
 \end{quote}
 \pagebreak
\item
Against manned space travel
\begin{quote}
{\it Astronauts are not effective in scientific research. For the
cost of taking astronauts safely to the Moon or planets and
bringing them back, one could send many hundreds of robots
that could do far more in the way of exploration.}
 \end{quote}
\begin{quote}
{\it The only technology for which the manned space flight program is well suited is the technology of keeping people alive in space}
 \end{quote}
\item
On being wrong
\begin{quote}
{\it Over the centuries the world has been greatly damaged by political and religious leaders who were sure they knew the truth, or behaved as if they did, and were able to pass on this certainty to their followers...my message to graduating students, for the worlds’ sake as well as for your own, is that when you go forth and get things wrong, as you inevitably will, that you be willing to recognize that you have been wrong, and even be a little proud that as scientists or engineers or architects you are able to know that you were wrong.}
 \end{quote}

\end{itemize}

\subsection{Other Facets}



Weinberg was an intellectual, widely read in history and poetry (he seemed to particularly enjoy Dylan Thomas), a devotee of classical music concerts and a master of the clear and precise explanation captured by a pithy phrase. But he also was a good citizen with strong beliefs about politics, economics and other topics in the public realm. He supported zionism and defended the state of Israel on many occasions, such as by cancelling trips to the UK in response to what he thought were unfair actions against Israel by British universities. 

He actively promoted science in education programmes in the US, fighting in particular against people who tried to suppress teaching Darwin's theory of evolution. He campaigned to promote the SSC, acting as a voice for the high energy physics community on Capitol Hill. He was also proud to have started and directed the Jerusalem Winter School, that has been successfully running for more than 40 years.

He consulted on defense issues and for some time participated in the group JASON (July, August, September, October, November), a group of top scientists that advised the US government on defense issues. He had a very strong work ethic and famously worked mostly at home watching old movies at the same time he performed long calculations.

During his later years he strongly opposed the open-carry laws that allowed university students in Texas to bring weapons to class. He continued to promote nuclear disarmament and to argue against the utility of sending humans to space.

He found he enjoyed being controversial on cultural issues like religion, philosophy, the science wars and so on, bringing to these discussions an articulate reasonableness backed up by a knowledgeable and logically structured mind. But he also recognised that it can sometimes be difficult to argue based on logical grounds if the other side does not adhere to logical thinking.

He was a family man, happily married for more than 65 years, loving his daughter and granddaughter and enjoying Valentine's Day dancing and dinner celebrations with his wife until his last years. His strong personality and booming New York accent combined with seemingly infinite knowledge and relentlessly logical thinking could at times be intimidating. But he was also a caring person who treasured human values and a sense of humour, often laughing loudly of his own jokes and at himself: 
{\it I know atheists that have the highest moral standards, for instance myself!}
 
A Renaissance Man who fell by chance into the modern world: few would wield so effectively such broad knowledge and fewer still would reach such pinnacles of achievement. His departure makes one feel that the bright light of intellect penetrates a bit less into the surrounding darkness. He is sorely missed.

\section*{Acknowledgements}
We thank Steven Weinberg for many things, but mostly for his integrity and for being such an excellent role model; for his students (like ourselves), for his postdocs and collaborators, and more broadly for theoretical physicists and for all scientists. His life-long commitment to science in all its aspects is a continuing source of inspiration, and likely will remain so for generations to come. We are grateful to our colleagues Daniel Baumann, Shanta de Alwis, Carlos Ord\'o\~nez and Sonia Pab\'an for sharing their insights and guidance while preparing this review, and we thank the Royal Society for the challenge of writing this essay and for their great patience for the many delays in its delivery.

\begin{appendix} 
\section{Honours and Awards}

Weinberg received a great many awards, honorary degrees and other accolades over a long and distinguished career. A select subset of these is listed here\footnote{For a more complete listing of his accolades see this \href{https://en.wikipedia.org/wiki/Steven_Weinberg\#Honors_and_awards}{link}.}

 \begin{itemize}
 \item The 1979 Nobel Prize in Physics. 
 
 \item  The foreign membership at the Royal Society 1981.
 
 \item The Heineman Prize for Mathematical Physics 1977. 
 
 \item  The National Medal of Science 1991. 
  
  \item The Benjamin Franklin Medal 2004.
  
  \item The James Joyce Medal 2009. 

   \item The Breakthrough Prize 2020. 
\end{itemize}

Given his broad and prolific career he also received as a scientific writer and humanist including:

\begin{itemize}
\item The Steel Foundation science writing award 1977.

\item The Lewis Thomas Award for writing about science, recognizing the scientist as a poet 1999. 

\item The Humanist of the year 2002 from the American Humanist Society.

\end{itemize}

\end{appendix}

\end{document}